\documentclass[preprint,onecolumn,a4paper,english,prb]{revtex4}
\usepackage{lmodern}

\usepackage[T1]{fontenc}
\usepackage[utf8]{inputenc}
\usepackage{textcomp}
\usepackage{amsmath}
\usepackage{graphicx}
\usepackage{epstopdf}
\usepackage{natbib} 
\usepackage{hyperref}
\usepackage{ulem}
\usepackage{amssymb}
\makeatletter

\@ifundefined{textcolor}{}
{%
 \definecolor{BLACK}{gray}{0}
 \definecolor{WHITE}{gray}{1}
 \definecolor{RED}{rgb}{1,0,0}
 \definecolor{GREEN}{rgb}{0,1,0}
 \definecolor{BLUE}{rgb}{0,0,1}
 \definecolor{CYAN}{cmyk}{1,0,0,0}
 \definecolor{MAGENTA}{cmyk}{0,1,0,0}
 \definecolor{YELLOW}{cmyk}{0,0,1,0}
 }

\usepackage{babel}

\begin{document}
\begin{abstract}

This is the version of the article before peer review or editing, as submitted by an author to Nanotechnology. IOP Publishing Ltd is not responsible for any errors or omissions in this version of the manuscript or any version derived from it. The Version of Record is available online at https://doi.org/10.1088/1361-6528/ab0450.

Amorphous aluminum oxide Al$_2$O$_3$ (a-Al$_2$O$_3$) layers grown by various deposition techniques contain a significant density of negative charges. In spite of several experimental and theoretical studies, the origin of these charges still remains unclear. We report the results of extensive Density Functional Theory (DFT) calculations of native defects - O and Al vacancies and interstitials, as well as H interstitial centers - in different charge states in both crystalline $\alpha$-Al$_2$O$_3$ and in a-Al$_2$O$_3$. The results demonstrate that both the charging process and the energy distribution of traps responsible for negative charging of a-Al$_2$O$_3$ films [M. B. Zahid et al., IEEE Trans. Electron Devices 57, 2907 (2010)] can be understood assuming that the negatively charged O$_{\textrm{i}}$ and V$_{\textrm{Al}}$ defects are nearly compensated by the positively charged H$_{\textrm{i}}$, V$_{\textrm{O}}$ and Al$_{\textrm{i}}$ defects in as prepared samples. Following electron injection, the states of Al$_{\textrm{i}}$, V$_{\textrm{O}}$ or H$_{\textrm{i}}$ in the band gap become occupied by electrons and sample becomes negatively charged. The optical excitation energies from these states into the oxide conduction band agree with the results of exhaustive photo-depopulation spectroscopy (EPDS) measurements [M. B. Zahid et al., IEEE Trans. Electron Devices 57, 2907 (2010)]. This new understanding of the origin of negative charging of a-Al$_2$O$_3$ films is important for further development of nanoelectronic devices and solar cells.
\end{abstract}

\title{The origin of negative charging in amorphous Al$_2$O$_3$ films: The role of native defects}

\author{Oliver A. Dicks}
\email[Corresponding author. Email address: ]{oliver.dicks.11@ucl.ac.uk}
\affiliation{Department of Physics and Astronomy, University College London, 
Gower Street, London WC1E 6BT, United Kingdom}

\author{Jonathon Cottom}
\affiliation{Department of Physics and Astronomy, University College London, 
Gower Street, London WC1E 6BT, United Kingdom}

\author{Alexander L. Shluger}
\affiliation{Department of Physics and Astronomy, University College London, 
Gower Street, London WC1E 6BT, United Kingdom}

\author{Valeri V. Afanas'ev}
\affiliation{Department of Physics and Astronomy, University of Leuven, Celestijnenlaan 200d, 3001 Heverlee, Belgium}

\maketitle

\section{Introduction}\label{defectintro}

Reliable characterization and identification of electron traps in thin insulating films is of utmost importance for eliminating or limiting the impact of these defects on the performance of electronic devices. In particular, it is has been known for a long time that amorphous aluminium oxide Al$_2$O$_3$ (a-Al$_2$O$_3$) layers grown using different deposition techniques contain a significant density of negative charges ~\cite{Afanasev2004,Govoreanu2006,Novikov2009,Zahid2010,Li2014} of still unclear origin. Specifically, understanding of electron trapping in amorphous alumina is important in the development of various nanoelectronic devices, including charge trap flash memory cells~\cite{Zahid2010,Li2014} and amorphous Indium Gallium Zinc Oxide (a-IGZO) transistors~\cite{Ok2015}. Furthermore, in some applications the presence of charge is desirable. For example, in silicon solar cells a-Al$_2$O$_3$ layers with a significant density of fixed negative charge are used to achieve electrostatic passivation ~\cite{Kessels_AlO_2012,Nemeth_PV_2017,Bonilla_PVpass_2017} by introducing substantial band bending at the silicon side of the Si/a-Al$_2$O$_3$ interface. This leads to a reduction of the surface recombination velocity thus improving the solar cell efficiency. 

It has been suggested that the formation of negative charges in alumina is result of electron transfer from silicon to energetically deep electron traps inside the oxide. A broad variety of models for the electron traps have been introduced, ranging from oxygen interstitials \cite{Kuhnhold_SiAlO_JAP2016} to oxygen vacancies \cite{Van_de_Walle_MEE_2013} and aluminum vacancies \cite{Kessels_AlO_2012}. Testing the validity of these models requires understanding of how these trapping site models can explain the thermally-activated increase of the negative charge observed when annealing a-Al$_2$O$_3$ layers on Si at temperatures below 500 $^{\circ}$C \cite{Kuhnhold_SiAlO_APL2016}. These temperatures are insufficient for defect generation since they are well below the temperature range needed for atomic re-arrangements in alumina, e.g., for crystallization which starts above 800 $^{\circ}$C\cite{Afanasev_APL_2002AlO}. The previous calculations~\cite{Dicks2017} suggest that intrinsic network sites do not provide energetically deep (>1 eV) trapping sites for electron polarons in a-Al$_2$O$_3$. Therefore in this work we investigate whether structural network imperfections, such as native defects and hydrogen ubiquitous in these samples, can explain the dominance of negative charging. 

The major challenge for atomic identification of electron trapping sites in a-Al$_2$O$_3$ concerns the absence of electron spin resonance (ESR) signals associated with these electron states. Despite several attempts to observe electrons trapped in a-Al$_2$O$_3$  using ESR, only signals stemming from its interface with the silicon substrate (silicon dangling bonds at the surface of the Si crystal or in the near-interface Si oxide layer) or with contaminants, mostly carbon-related, have been detected so far \cite{Stesmans_JoP_2001,Stesmans_JVSTB_2002,Stesmans_APL_2002,Stesmans_APL_2004,Stesmans_JAP_2005,Kessels_AlO_2012,Kuhnhold_SiAlO_JAP2016}. We note that crystallization of alumina as a result of high temperature annealing does not eliminate electron trapping sites, suggesting that they are not caused by disorder as in a-HfO$_2$\cite{Strand2017b}. Moreover, experiments consistently indicate that the defects in question are abundant in a-Al$_2$O$_3$ films. For example, measurements of the threshold voltage shifts in a-IGZO thin film transistors provide an estimate of the electron trap density of 1$\times 10^{13}/$cm$^2$ in a 30 nm a-Al$_2$O$_3$ `charge trapping' layer~\cite{Li2014}. 

\begin{figure}
\centering
\includegraphics[width=0.8\linewidth]{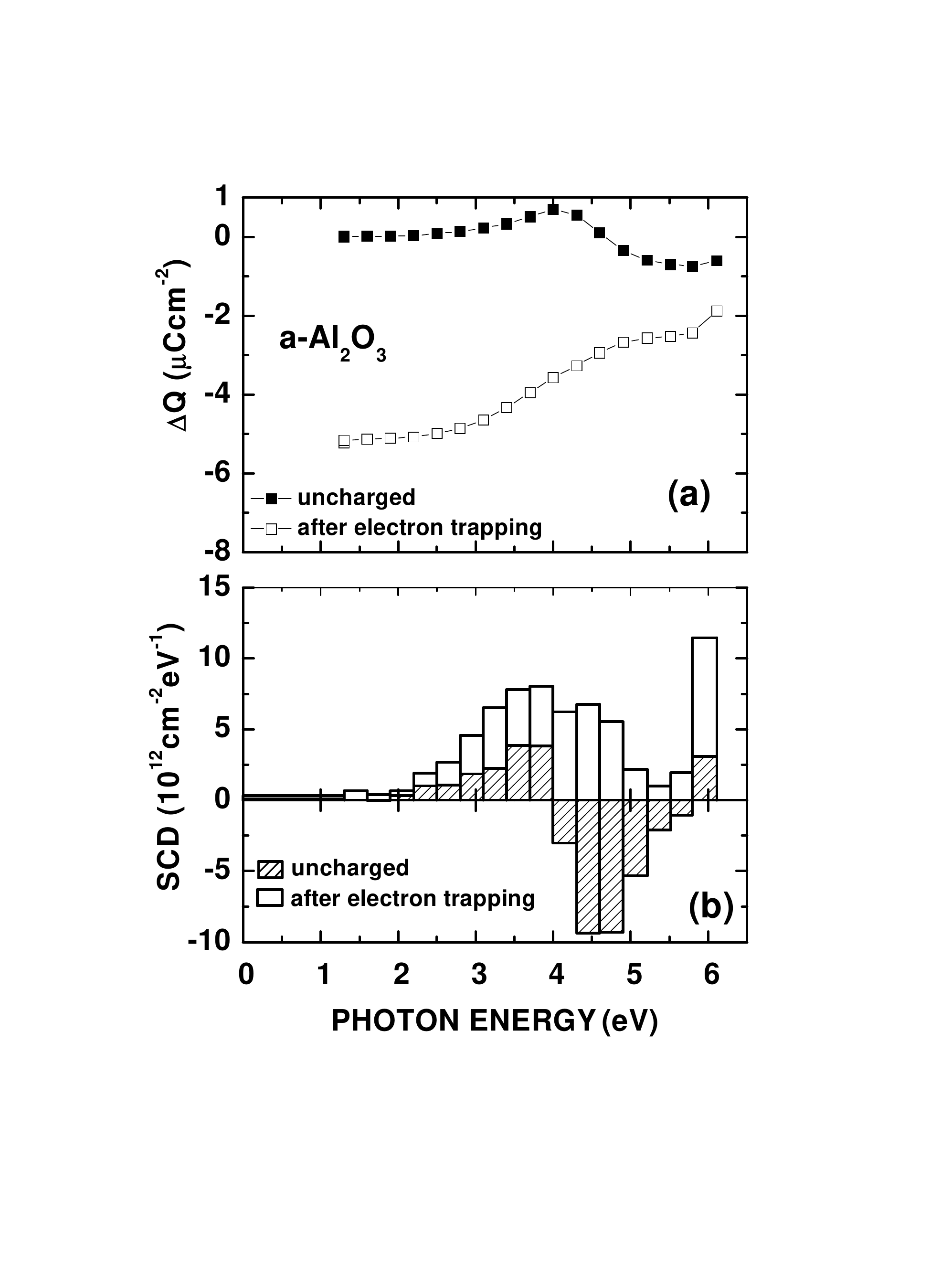}
\caption{\label{fig:PDS} Oxide charge density variations induced by illumination of Si/SiO$_2$(4 nm)/Al$_2$O$_3$(20 nm) stacks (a) and the inferred spectral charge density diagrams (b) for uncharged and electron injected a-Al$_2$O$_3$ films. Negative SCD values correspond to the net electron trapping caused by the electron photoemission from silicon into the tunneling SiO$_2$ layer. The inferred energy distribution of the electron traps with respect to the CBM are shown for the as-grown (neutral) films (dashed bars) and after injecting electrons resulting in negative charging of the film (open bars).}
\label{single}
\end{figure}

Whilst ESR measurements have proved challenging, the analysis of electron trapping in Al$_2$O$_3$ thin films in ref. \cite{Zahid2010} has provided the trap density at various depths within the film, and the position of their energy levels below the conduction band minimum (CBM) using exhaustive photo-depopulation spectroscopy (EPDS). EPDS measures the energy levels of the defects by monitoring the density of electronic charge remaining in the film after the photo-excitation of electrons into the Al$_2$O$_3$ conduction band \cite{Afanas'ev_2007_IPErev,Zahid2010,Strand2017b}. Charging of the Al$_2$O$_3$ was performed by electron tunneling through a 4-nm SiO$_2$ tunneling layer at high positive voltage. After charging, the sample was left for several hours in darkness, allowing all shallow traps and traps close to the gate to discharge. After significant charging of the alumina films by electron injection from silicon, EPDS measurements revealed a broad range of electron trap energy levels 2-4 eV below the CBM (see Fig. \ref{fig:PDS}) with a trap areal density well above $1\times 10^{13}/$cm$^2$ \cite{afanas2014invited}. Furthermore, as can be seen in Fig. \ref{fig:PDS}, the EPDS trap spectra measured prior to electron injection in the alumina film reveal a similar energy distribution of the occupied electron states in the oxide bandgap, albeit with a much lower density. This observation suggests that the fixed negative charges commonly encountered in the as-deposited alumina layers are related to the partial filling of electron traps already present during Al$_2$O$_3$ film synthesis.

The EPDS measurements were confirmed by gate side trap spectroscopy when injecting electrons from silicon and sensing' (GS-TSCIS) measurements which show a peak defect density of $1.6\times 10^{19}/$cm$^3$ at approximately 3.4 eV below the Al$_2$O$_3$ CBM, with a significant distribution of traps from ~3.0 eV below the CBM~\cite{Zahid2010}. In addition, a band of shallow traps was also identified at 1.6-1.8 eV below the CBM, with a trap density of approximately $1.6\times 10^{19}/$cm$^3$ in the bulk of the Al$_2$O$_3$ films~\cite{Zahid2010}.

To elucidate the origin of negative charging and the nature of trapping sites responsible for the measured energy spectrum of trapped electrons in the alumina bandgap, shown in Fig. \ref{fig:PDS}, we performed Density Functional Theory (DFT) calculations. We assumed that a-Al$_2$O$_3$ films contain native defects which are generated during the film deposition process, where the structure deviates from the stoichiometric `pure' amorphous topology, similar to intrinsic and extrinsic defects in crystalline $\alpha$-Al$_2$O$_3$ which have been widely studied computationally~\cite{Peacock2003,Weber2011,Holder2013,Gordon2014,Li2014a}. We compare the structural and electronic properties of interstitial hydrogen (H$_{\textrm{i}}$), oxygen vacancies (V$_{\textrm{O}}$), oxygen interstitials (O$_{\textrm{i}}$), aluminum vacancies (V$_{\textrm{Al}}$) and aluminum interstitials (Al$_{\textrm{i}}$) in a-Al$_2$O$_3$ and $\alpha$-Al$_2$O$_3$. The calculated charge transition levels and Kohn-Sham (KS) energy levels of the defects are compared to experimental values~\cite{Zahid2010} to enable identification of the defects responsible for the negative charging of a-Al$_2$O$_3$ films. We conclude that negative charging of a-Al$_2$O$_3$ films is caused predominantly by electron trapping by native defects – O  vacancies and Al interstitial ions, as well as by H interstitials. These positively charged defects are compensated in as prepared samples by negatively charged O interstitial ions and Al vacancies. The charge balance is shifted to negative charge as a result of electron trapping which also creates the occupied states in the gap observed experimentally. 

\section{Computational Methodology}\label{sec:methodology}

The methods and potentials employed to model the a-Al$_2$O$_3$ systems have a significant impact on the structures produced. In experiment the choice of substrate and growth conditions have a large effect on the density of the a-Al$_2$O$_3$ films~\cite{Lamparter1997,Gutierrez2002,Choudhary2015}. The systems considered in this paper were characterized in detail in the previous study\cite{Dicks2017}, with their densities being representative of the ALD films that are most relevant for device applications. 

A range of defects in a-Al$_2$O$_3$ were investigated using 10 amorphous structures generated using molecular dynamics and a melt-quench method, described previously~\cite{Dicks2017}. These calculations were performed using the LAMMPS code~\cite{Plimpton1995}, 360 atom cells, and a Born-Mayer type inter-atomic potential that had been previously used to generate a-Al$_2$O$_3$~\cite{Blackberg2014,Gutierrez2002}. Briefly, the melt-quench procedure works as follows. The 360 atom cells of $\alpha$-Al$_2$O$_3$ used as the initial structure were equilibrated at 300K for 10 ps, and then heated to 5000K over 20 ps and equilibrated in the melt. The structures were then cooled to 1K at a cooling rate of 10Kps$^{-1}$. The NPT ensemble was used with a time step of 0.1 fs. The cell geometry was then re-optimized using DFT. 

This method produced a-Al$_2$O$_3$ structures with an average density of 3.14 gcm$^{-3}$, in agreement with experimental measurements of 2.97-3.20 gcm$^{-3}$~\cite{Groner2004,Ilic2010,Ok2015}. As discussed in detail in ref. \cite{Dicks2017}, the obtained structural properties agree well with the experimentally measured distribution of coordination numbers~\cite{Lee2009} and radial distribution functions from x-ray and neutron diffraction studies~\cite{Lamparter1997}. In particular, 53\%, of Al atoms are 4-coordinated by O atoms, whereas 37\% and 10\% are 5- and 6-coordinated, respectively.  This shows that in amorphous films most Al atoms are under-coordinated with respect to $\alpha$-Al$_2$O$_3$, where all Al atoms are 6-coordinated with O.

The calculations of the electronic structure were performed using the CP2K package~\cite{VandeVondele2005}. The PBE0-TC-LRC~\cite{Guidon2009} functional was used with a cutoff radius of 3.0 $\text{\AA}$, tuned to minimize deviation from straight line error and to ensure that Koopmans' condition is obeyed (as described in \cite{Dicks2017}) similar to the implementation in \cite{Kronik2012,Karolewski2013}. The DZVP-MOLOPT-SR-GTH~\cite{VandeVondele2007} basis sets were used for O, Al and H with the Goedecker-Teter-Hutter (GTH) pseudopotentials~\cite{Goedecker1996,Hartwigsen1998}, and the auxiliary density matrix method~\cite{Guidon2009} (ADMM) was used to speed up the calculations. The plane wave energy cutoff was set to 500 Ry and the SCF convergence criterion was 10$^{-6}$ a.u.. The calculated Kohn-Sham (K-S) single-electron band gap in $\alpha$-Al$_2$O$_3$ is 8.6 eV, close to the experimental optical band gap of 8.7 eV. The average value of the K-S band gap calculated from ten structures in a-Al$_2$O$_3$ is 5.5 eV. 

We note that there is a marked shift both in the measured and calculated band gap between the crystalline and the amorphous phases, with $\alpha$-Al$_2$O$_3$ having a band gap of 8.7 eV while the range of band gaps observed for a-Al$_2$O$_3$ grown using ALD is 6.0 eV - 7.0 eV \cite{AfanasEv2011,Filatova2015,Momida2006}, this represents a significant reduction of 20 - 30 \%. This phenomenon and its relationship to the film density, structure, and dielectric properties have been examined theoretically in~\cite{Momida2006} and experimentally by several groups using a variety of techniques.\cite{Filatova2015,SNIJDERS200297,Afanasev_APL_2002AlO,PhysRevB.83.094201} While there is not yet a clear consensus as to the exact nature of this band gap reduction, it is clear that the hybridization and the relative position of both the valence band and conduction bands are affected. \cite{Momida2006,Colleoni2015,Filatova2015,PhysRevB.83.094201} This last point is of critical importance when considering defects in these structures and the associated states in the band gap. Any shift in the band edges may result in the same defect behaving dramatically differently in the crystal and the amorphous phases. A deep state in the crystal can transform into shallow state in the amorphous film and some shallow states can be removed from the band gap all together. The full extent of this electronic structure, density, and motif dependence will be explored in a future publication, and as such is beyond the scope of this work.

Defects were then created in the 360 atoms cells of both a-Al$_2$O$_3$ and $\alpha$-Al$_2$O$_3$ by the addition or removal of an ion at a variety of sites, in order to sample the effect of different densities, band gaps and local atom coordination numbers. The structures were then allowed to undergo full geometry optimization using the PBE0-TC-LRC~\cite{Guidon2009} hybrid functional, with a maximum force convergence criteria of 0.05 eV/\text{\AA}.

Unlike in crystals, where there is a finite number of non-equivalent defect sites within the primitive unit cell, in amorphous structures the lack of periodicity means no sites are exactly equivalent, and so any defect properties will have a wide distribution of values, dependent on local geometric structure and the bulk properties of the specific amorphous cell. Full sampling of different local configurations even within 360 atom cells using the non-local density functional is beyond our computational means. The aim of this work is to ascertain which defects are likely to be responsible for negative charging of a-Al$_2$O$_3$ films and estimate the range of corresponding formation energies and gap states. To achieve that, we calculate the defect properties in ten statistically independent cells obtained using the method described above and in refs.~\cite{Kaviani2017,El-Sayed2015}. This allows a variety of local defect geometries to be sampled, which have different bond lengths and local coordination numbers, whilst also allowing for variation in bulk properties that can affect the stability of certain defect configurations and charge states. To create vacancies, atoms are removed at random, but so the distribution of the atom coordination numbers matches that of the distribution across the 10 cell geometries. Interstitial defects are created by adding an atom at a random position in the cell, with a consideration of minimum inter-atomic distances, and then performing a geometry relaxation. 

\subsection{Defect formation energies}

The defect formation energies and charge transition levels for defects in crystalline $\alpha$-Al$_2$O$_3$  are calculated using the method described in ref. \cite{Freysoldt2014}. The chemical potential of hydrogen used in the calculation of the formation energies is $\mu_{\textrm{H}} = E_{\textrm{H}_{2}}/2$, where E$_{\textrm{H}_{2}}$ is the total energy of an H$_2$ molecule. The chemical potential of oxygen in O-rich conditions was taken to be $\mu^{0}_{\textrm{O}} = E_{\textrm{O}_{2}}/2$, where E$_{\textrm{O}_{2}}$ is the total energy of an O$_2$ molecule in the triplet state.  Although DFT is known to over-bind the O$_2$ molecule leading to over estimation of the dissociation energy~\cite{Klupfel2012}, this has not been corrected in this paper. It is known that generalized gradient approximation functionals over estimate the dissociation energy by large amounts (>1 eV)~\cite{Klupfel2012}, but hybrid functionals, such as PBE0, better correct the self interaction error, leading to much smaller errors of approximately 0.3 eV~\cite{Klupfel2012}. Using this method allows us to compare with other papers, and, as the main properties of interest are the charge transition levels and K-S one electron energy levels, the chemical potential will have no effect on these properties.

The Al reference chemical potential, $\mu^{0}_{\textrm{Al}}$, was calculated from a 256 atom cell of bulk Al in the cubic phase, after a full cell relaxation, where $\mu^{0}_{\textrm{Al}} = E_{\textrm{Al}}$. The DFT calculation used the same PBE0-TC-LRC~\cite{Guidon2009} functional and DZVP-MOLOPT-SR-GTH~\cite{VandeVondele2007} basis sets as the other calculations in this paper. This corresponds to the Al-rich condition.

The chemical potentials of various growth conditions can then be set by adjusting the $\Delta\mu_{\alpha}$ parameters, which correspond to differences in the chemical potentials from their reference values ($\mu^{0}_{\alpha}$), as they must obey certain conditions. First the sum of the chemical potentials must equal the formation enthalpy of alumina ($\Delta H_{\textrm{Al}_2\textrm{O}_3}$),

\begin{equation}
2\Delta\mu_{\textrm{Al}}+3\Delta\mu_{\textrm{O}} = \Delta H_{\textrm{Al}_2\textrm{O}_3}.
\end{equation}

They must also obey the conditions where
\begin{equation}
\Delta\mu_{\textrm{Al}} < 0
\end{equation}
and,
\begin{equation}
\Delta\mu_{\textrm{O}} < 0
\end{equation}
in order to prevent them reverting to their pure elemental form. If these conditions are obeyed then O-rich conditions correspond to when $\Delta\mu_{\textrm{O}} = 0$, and Al-rich conditions when $\Delta\mu_{\textrm{Al}} = 0$.

Both a-Al$_2$O$_3$ films grown experimentally and our theoretical models are metastable and never in thermodynamic equilibrium. It is likely that the creation of defects is a dynamic process that occurs during film deposition, and cannot be described by Boltzmann statistics. Nevertheless we use the same method to estimate the charge transition levels of defects in different positions in amorphous structures. This provides comparison with these energies in the crystalline phase, and some guidance for expected defect charge states before and after electron injection. The average Gibbs free energy of formation of defects in a-Al$_2$O$_3$ is given in the supplementary material~\cite{Supplementary}.

Charge corrections to the energy, as a result of the interactions between point charges in periodic supercells~\cite{Lany2008,Komsa2012} were included in the calculations of the defect formation energies. The corrections led to small shifts in the positions of the charge transfer levels by no more than 0.4 eV, due to the high dielectric constant of amorphous Al$_2$O$_3$ (9.6~\cite{Shamala2004}) and the large size of the cells.

\section{Results of Calculations}

\subsection{Interstitial hydrogen}

Due to its presence in almost every growth environment, hydrogen is a common impurity in most metal oxides and semiconductors, including Al$_2$O$_3$~\cite{El-Aiat1982,Groner2004,Jennison2004}. Therefore it is not surprising that the negative charging observed experimentally~\cite{Jennison2004,Li2014} in a-Al$_2$O$_3$ thin films has been attributed to interstitial hydrogen (H$_{\textrm{i}}$). A tunneling conductivity via a mid-band gap state was observed to increase after an increase of the H content in alumina films~\cite{Jennison2004}. DFT studies of H$_{\textrm{i}}$ in crystalline Al$_2$O$_3$, using LDA~\cite{Peacock2003}, GGA~\cite{Jennison2004} and hybrid~\cite{Holder2013,Gordon2014,Li2014a} functionals, confirm that interstitial hydrogen has a mid-band gap energy level in alumina, close to its band offset with Si. Due to the lack of computational data on interstitial hydrogen in amorphous alumina, DFT studies of crystalline Al$_2$O$_3$ are used for comparison in this section.

To create the defects, neutral hydrogen atoms were inserted at random positions within the 10 amorphous Al$_2$O$_3$ geometries, while ensuring initial O-H distances were greater than 1.6 $\textrm{\AA}$, and then allowed to relax. This allowed the H to be positioned close to O ions with a range of coordination numbers during relaxation. After calculating the properties of the H$_{\textrm{i}}^{0}$, the different charge states were investigated by the addition or removal of an electron to the system followed by a full geometry optimization. 

\subsubsection{H$_{\textrm{i}}^{+}$}

\begin{figure}
\centering
\includegraphics[width=1.0\linewidth]{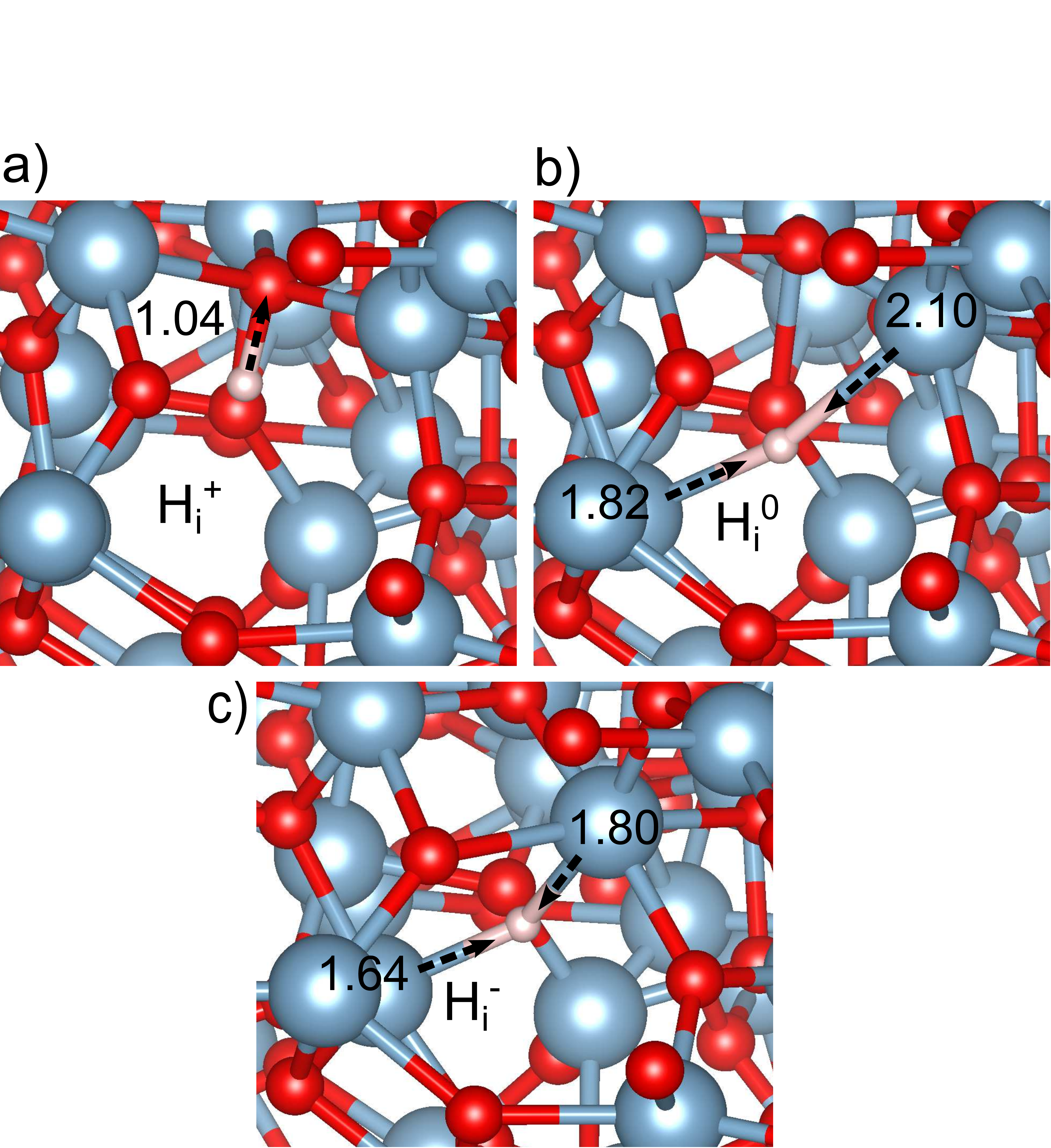}
\caption{\label{fig:Hipos} (Color version online) The structural configurations of the 3 H$_{i}$ charge states, similar to those seen in crystalline Al$_2$O$_3$, with Al colored grey (large sphere), O red (medium sphere) and H white (small sphere). The arrows show the direction of relaxation, with the labels showing the length of the bond. a) the H$_{\textrm{i}}^{+}$ defect which forms an OH bond. b) H$_{\textrm{i}}^{0}$ which lies between 2 Al ions as atomic hydrogen. c) H$_{\textrm{i}}^{-}$ also sits between 2 Al ions, though they relax towards the negatively charged ion.}
\label{single}
\end{figure}

In crystalline alumina there are three possible charge states of interstitial hydrogen, each of which has different structural and bonding characteristics. Previous DFT calculations predict that H$_{\textrm{i}}^{+}$ forms an OH bond with a nearest neighbor oxygen in $\alpha$-Al$_2$O$_3$~\cite{Holder2013,Gordon2014,Li2014a} and $\theta$-Al$_2$O$_3$~\cite{Li2014a}, with O-H bond lengths of approximately 1.0 and 1.1 $\textrm{\AA}$ respectively. As can be seen in Fig \ref{fig:Hipos}a, the H$_{\textrm{i}}^{+}$ defect demonstrates similar behavior in a-Al$_2$O$_3$, with the proton forming an OH bond with a nearest neighbor oxygen. The average O-H bond length over the 10 amorphous samples is 1.00 $\textrm{\AA}$, with a range of 0.96-1.07 $\textrm{\AA}$.  

Out of the 10 H$_{\textrm{i}}^{+}$ configurations in a-Al$_2$O$_3$, 7 formed an OH bond with a 2-coordinated O ion ($^{[2]}$O), and 3 with a 3-coordinated O ($^{[3]}$O). The under-coordination of the O ions means that no Al-O bonds have to be broken to form the lowest energy configurations of the defect. This can be compared to hybrid functional calculations of $\alpha$-Al$_2$O$_3$~\cite{Li2014a}, where 2 of the 4 oxygen ion's Al-O bonds are broken to form the OH configuration, whilst in $\theta$-Al$_2$O$_3$ the proton bonds with a 3-coordinated O and no O-Al bonds are broken. The H$_{\textrm{i}}^{+}$ defects in a-Al$_2$O$_3$ where the OH bond includes a $^{[2]}$O have an average formation energy that is 0.9 eV lower than that with a $^{[3]}$O. This could be due to the fact that the addition of a proton significantly  lowers the energy of the localized O $\sigma_{2p}^{*}$ type orbitals observed at the top of the valence band in bulk a-Al$_2$O$_3$, which are a direct result of the O under-coordination.

\subsubsection{H$_{\textrm{i}}^{-}$}

In 7 out of the 10 interstitial hydrogen defects calculated, the negatively charged hydrogen interstitial forms an isolated H$^{-}$ ion, as shown in Fig. \ref{fig:Hipos}c. This is similar to the structural geometry of the defect observed in $\alpha$-Al$_2$O$_3$ and $\theta$-Al$_2$O$_3$~\cite{Li2014a}. In the other 3 configurations an [H$_{\textrm{i}}^{+} + 2e_{\textrm{CBM}}$] defect is formed.

In the formation of the H$_{\textrm{i}}^{-}$ defect, the electron that is introduced to the system localizes on the hydrogen, giving the ion an average Bader charge of -0.9$|e|$. The negative charge of the hydrogen ion causes significant relaxation of the 2 nearest neighbor Al ions, which are attracted to the interstitial hydrogen due to their formal +3 positive charge. An example of this relaxation can be seen in Fig. \ref{fig:Hipos}. During this relaxation the Al-H bond contracts by 0.19 to 0.36 $\textrm{\AA}$, when compared to the bond lengths in the neutral charge state. Similar relaxation of the 2 nearest neighbor Al towards the H$_{\textrm{i}}^{-}$ defect is observed in crystalline Al$_2$O$_3$~\cite{Gordon2014,Li2014a}, with the hydrogen sitting equidistant between the Al ions with a bond length of 1.67 \AA. Relaxation of multiple cations towards the defect also occurs in amorphous HfO$_2$~\cite{Kaviani2017}, and crystalline MgO and La$_2$O$_3$~\cite{Li2014a}. This differs from the relaxation observed in more covalent oxides, such as amorphous SiO$_2$ ~\cite{El-Sayed2015} and crystalline SiO$_2$ and GeO$_2$~\cite{Li2014a}, where the negatively charged hydrogen bonds to a single cation, causing large relaxation of the surrounding oxygens. This would suggest that the ionicity of the material strongly affects the configuration and bonding character of the H$_{\textrm{i}}^{-}$ defect.

\subsubsection{H$_{\textrm{i}}^{0}$ and the [H$_{\textrm{i}}^{+} + e_{\textrm{CBM}}$] defect}

The neutral hydrogen interstitial, H$_{\textrm{i}}^{0}$, is a metastable defect in alumina. In crystalline $\alpha$-Al$_2$O$_3$ H$_{\textrm{i}}^{0}$ behaves like an isolated hydrogen atom and is 2 coordinated with Al~\cite{Li2014a}. It does not bind to any host atom and causes minimal relaxation of the surrounding lattice due to the charge neutrality~\cite{Li2014a}. The H$_{\textrm{i}}^{0}$ defect also exhibits similar characteristics in some locations in a-Al$_2$O$_3$. These are created by injecting an electron into the system containing the OH bond of the H$_{\textrm{i}}^{+}$ - as a result, this bond is broken and the electron localizes on the proton, forming an isolated hydrogen atom (see Fig. \ref{fig:Hipos}b). 

However, the neutral hydrogen interstitial defect in a-Al$_2$O$_3$ can also differ significantly from that in the crystalline material. In a-Al$_2$O$_3$ there are two possible configurations of the H$_{\textrm{i}}^{0}$ and, occasionally, the H$_{\textrm{i}}^{-}$ defects. An H atom or H$^{-}$ ion  introduced into a-Al$_2$O$_3$, can donate its electron(s) into the conduction band whilst forming an OH bond with a nearest neighbor oxygen (similar to the configuration seen in  Fig \ref{fig:Hipos}a). This can be termed an [H$_{\textrm{i}}^{+} + e_{\textrm{CBM}}$] or an [H$_{\textrm{i}}^{+} + 2e_{\textrm{CBM}}$] defect, where $e_{\textrm{CBM}}$ denotes a delocalized electron in the conduction band. DFT calculations using LDA functionals observed that for some oxides the H(+/-) charge transition level lies above the CBM~\cite{VanDeWalle2006}. This is attributed to a universal alignment of the H(+/-) energy level at approximately 4.5 eV below the vacuum level ~\cite{VanDeWalle2006}, meaning in oxides with a larger electron affinity, the energy needed to break the OH bond is greater than that required to place an electron in the conduction band. 

In a review of hydrogen and muonium data~\cite{Cox2003}, and their role as shallow or deep donors and acceptors, it is predicted that H$_{\textrm{i}}^{0}$ in materials with electron affinities greater than ~4 eV will auto-ionize and donate an electron into the conduction band, which is confirmed by ESR data~\cite{Cox2003}. Later studies, using HSE06, demonstrate that crystalline TiO$_2$ and SnO$_2$ show similar behavior~\cite{Li2014a}, both of which have electron affinities greater than 4 eV~\cite{Cox2003} and band gaps smaller than 5 eV~\cite{Li2014a}. Amorphous HfO$_2$ also has a [H$_{\textrm{i}}^{+} + e_{\textrm{CBM}}$] like defect, though the donated electron localizes at an intrinsic trap site, rather than delocalizing in the network at the conduction band minimum. This lowers its formation energy and makes this configuration energetically more favorable~\cite{Kaviani2017}. The trapping energy of an electron polaron in a-HfO$_2$ is approximately 1 eV, and so this compensates for the higher band gap of approximately 6.0 eV~\cite{Kaviani2017} (and a lower electron affinity of approximately 3 eV~\cite{Cox2003}). 

The [H$_{\textrm{i}}^{+} + e_{\textrm{CBM}}$] configuration is stabilized in  a-Al$_2$O$_3$ due to the lowering of the conduction band compared to $\alpha$-Al$_2$O$_3$ (see e.g. \cite{Filatova2015}). The lowering of the conduction band corresponds to an increase in the electron affinity towards the 4 eV suggested as the H(+/-) alignment level, and when the local structural relaxation lowers the energy surrounding the defect sufficiently, [H$_{\textrm{i}}^{+} + e_{\textrm{CBM}}$] can be more energetically stable than H$_{\textrm{i}}^{0}$ in the amorphous material. The [H$_{\textrm{i}}^{+} + e_{\textrm{CBM}}$] defect can be considered a meta-stable state in a-Al$_2$O$_3$, which is not observed in the crystalline material where the conduction band is higher, and the energy gained from structural relaxations is smaller. 

\subsubsection{Charge transition levels and energy levels of H$_{\textrm{i}}$}

\begin{figure}
\centering
\includegraphics[width=0.8\linewidth]{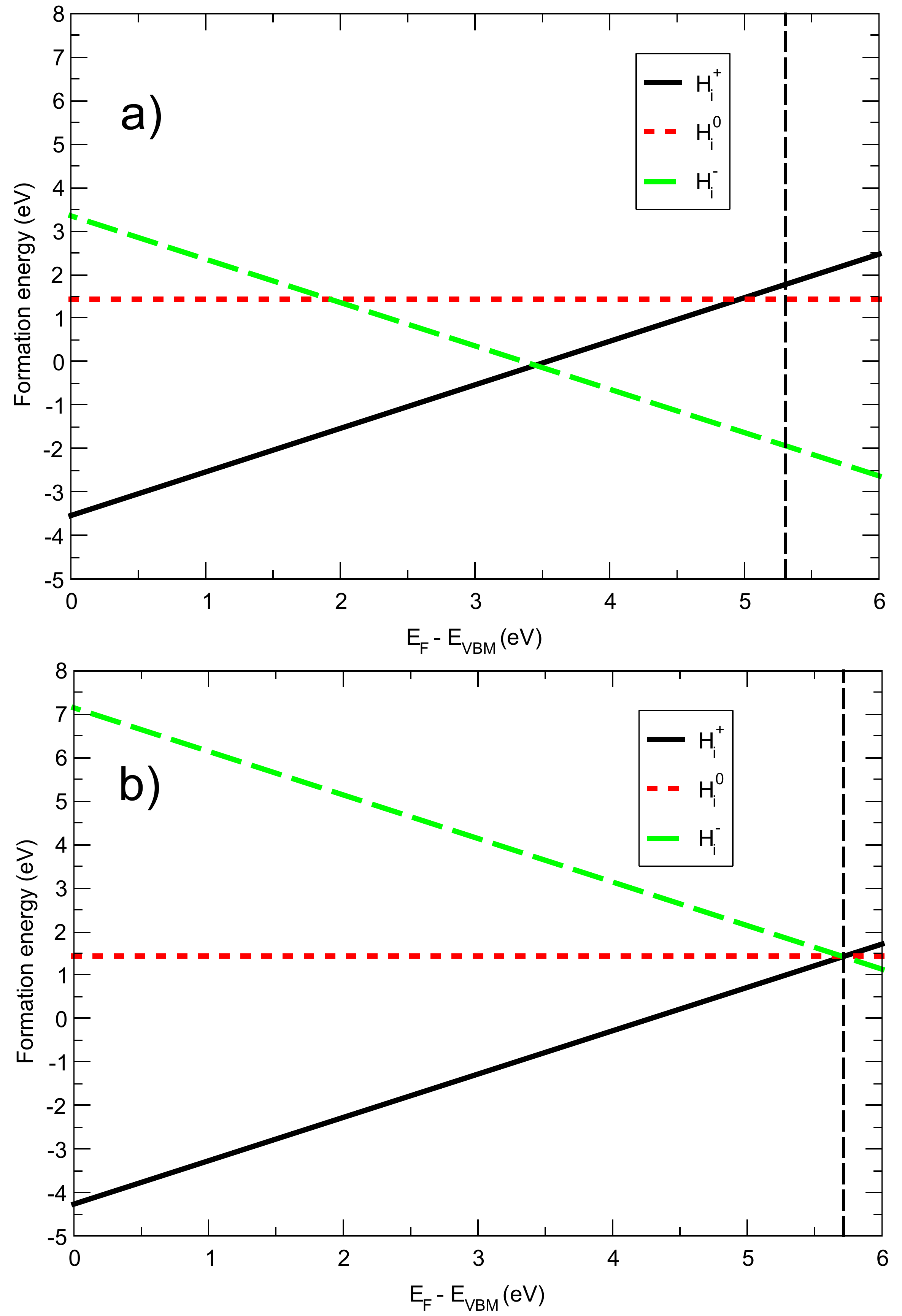}
\caption{\label{fig:HiEform} The formation energy of the different charge states of interstitial hydrogen against the Fermi energy with respect to the valence band. The dashed vertical line shows the position of the CBM. a) The formation energy of the H$_{\textrm{i}}^{+}$, H$_{\textrm{i}}^{0}$ and H$_{\textrm{i}}^{-}$ where the OH bond is broken in the neutral and negative charge state, as seen in Fig. \ref{fig:Hipos}b and Fig. \ref{fig:Hipos}c. b)  The formation energy of H$_{\textrm{i}}^{+}$, [H$_{\textrm{i}}^{+} + e_{\textrm{CBM}}$] and [H$_{\textrm{i}}^{+} + 2e_{\textrm{CBM}}$], here the (+/-) charge transfer level is at the CBM.}
\label{single}
\end{figure}
  
It is widely accepted that interstitial hydrogen in crystalline Al$_2$O$_3$ exhibits the so-called `negative-U' behavior, meaning its +1 (H$_{\textrm{i}}^{+}$) or -1 (H$_{\textrm{i}}^{-}$) charge states are more thermodynamically stable than its neutral state (H$_{\textrm{i}}^{0}$) for all values of the Fermi energy~\cite{Peacock2003,Holder2013,Gordon2014,Li2014a}. The calculations presented here show that this holds true for interstitial hydrogen in a-Al$_2$O$_3$. As can be seen from Fig. \ref{fig:HiEform}, there is only a (+/-) charge transition level for the H$_{\textrm{i}}^{-}$ defect, and the [H$_{\textrm{i}}^{+} + 2e_{\textrm{CBM}}$] defect necessarily has the charge transition level when the Fermi energy is aligned with the CBM. The formation energy of the neutral defect is never lower than that of H$_{\textrm{i}}^{+}$ or H$_{\textrm{i}}^{-}$ at any value of the Fermi energy.
  
\begin{figure}
\centering
\includegraphics[width=1.0\linewidth]{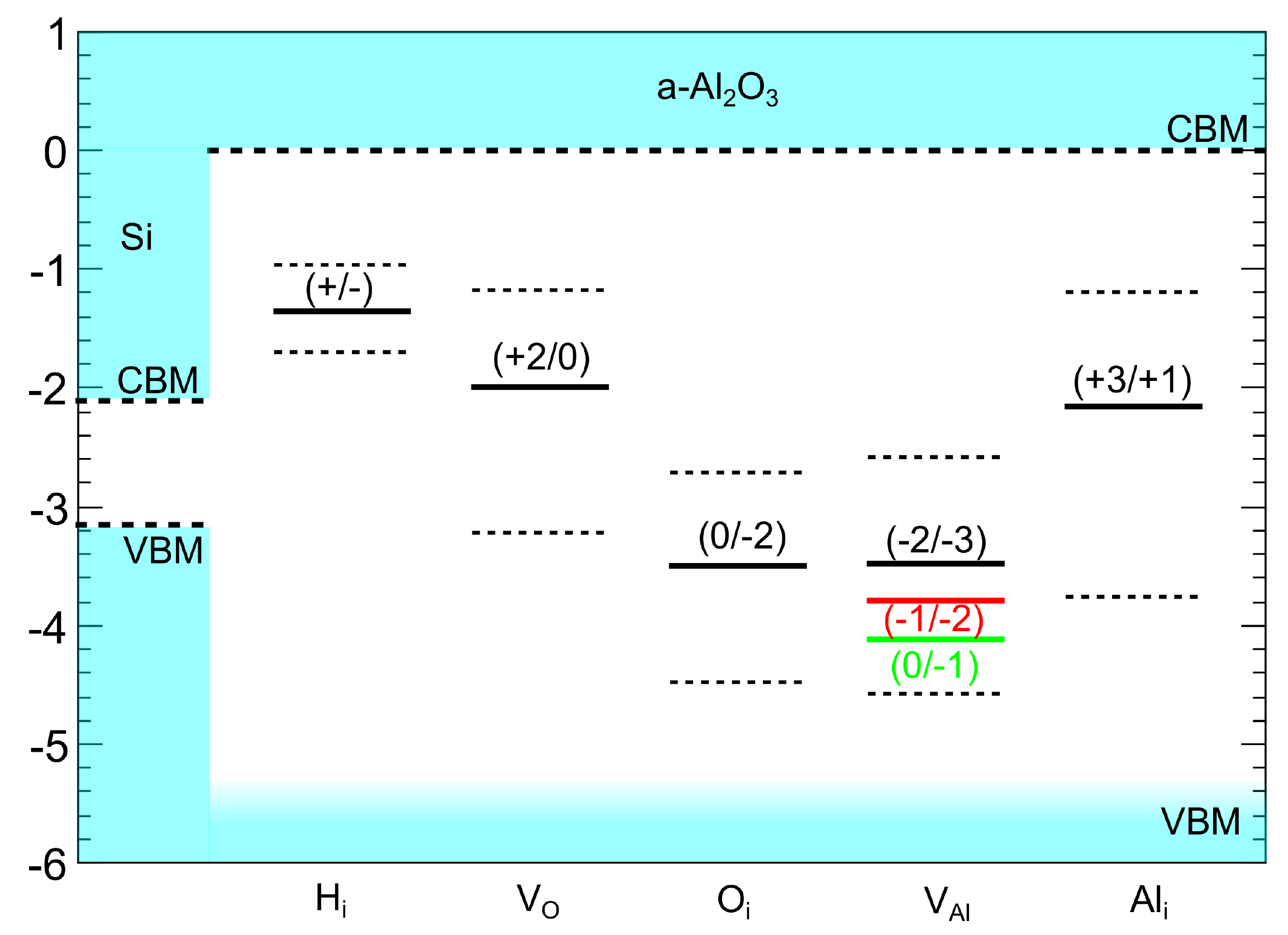}
\caption{\label{fig:qtrans} The average (solid lines) charge transition levels of H$_{\textrm{i}}$, V$_{\textrm{O}}$, O$_{\textrm{i}}$, V$_{\textrm{Al}}$ and Al$_{\textrm{i}}$ with respect to the a-Al$_2$O$_3$ CBM, the energy axis is in eV. The dashed lines represent the calculated range of the levels in the different samples. The Si/a-Al$_2$O$_3$ conduction band offset is taken from internal electron photoemission measurements~\cite{Afanasev2003}. }
\label{single}
\end{figure}
  
When only including the 7 H$_{\textrm{i}}^{-}$ configurations which have a charge transition level in the band gap (discounting those where H$_{\textrm{i}}^{+} + 2e_{\textrm{CBM}}$ form), the average (+/-) charge transition level lies 1.43 eV below the CBM, as can be seen in Fig. \ref{fig:qtrans}. The charge transition levels are shown with respect to the CBM in order to compare with experimentally measured conduction band offsets of a-Al$_2$O$_3$ with Si~\cite{AfanasEv2011}. The position of the level compared with the Si band offset suggests that the average (+/-) level lies approximately 1 eV higher than would be expected from the universal alignment of hydrogen levels in semiconductors predicted by Van de Walle and Neugebauer~\cite{VanDeWalle2006} using LDA functionals. However later studies~\cite{Li2014a} using hybrid functionals show that the (+/-) level of hydrogen interstitials in oxides can vary by more than $\pm$ 1 eV from the universal level in crystalline oxides. The charge transition level for a-Al$_2$O$_3$ calculated here is in good agreement with the one predicted for $\theta$-alumina~\cite{Li2014a}, which has a similar band gap. The position of the (+/-) level closer to the conduction band may be due to the strength of the OH bond of the H$_{\textrm{i}}^{+}$ defect. As no Al-O bonds have to be broken to form the OH bond in the amorphous material, the formation energy may be lowered compared to that in crystalline alumina. The H$_{\textrm{i}}^{-}$ defect in a-Al$_2$O$_3$, despite the slightly larger relaxations of the surrounding Al ions, more closely resembles the H$_{\textrm{i}}^{-}$ defect in crystalline alumina, resulting in a smaller decrease in formation energy and thus a higher lying (+/-) charge transition level.

In Fig.\ref{fig:qtrans} it can be seen that the (+/-) charge transition level lies above the Si conduction band, but only by an average of 0.6 eV, with the lowest (+/-) level within 0.2 eV of the Si CBM. As reported in the study by Zahid et al.~\cite{Zahid2010}, the negative charge traps in a-Al$_2$O$_3$ are populated via the tunneling mechanism. By varying the charging potential, traps at a range of energy levels in the alumina gap are populated when they are aligned with the injection level. Therefore H$_{\textrm{i}}^{-}$ could be responsible for the negative charging observed~\cite{Zahid2010}. However, the K-S energy levels of H$_{\textrm{i}}^{-}$ lie on average 4.8 eV below the CBM, approximately 1 eV lower than the extreme range of the trap levels measured by EPDS in ref.~\cite{Zahid2010}. So whilst it is possible for interstitial hydrogen to trap electrons in a-Al$_2$O$_3$~\cite{Govoreanu2006,Zahid2010,Li2014}, it is unlikely to be responsible for all the defect states observed by the EPDS measurements~\cite{Zahid2010}.
  
\begin{figure}
\centering
\includegraphics[width=1.0\linewidth]{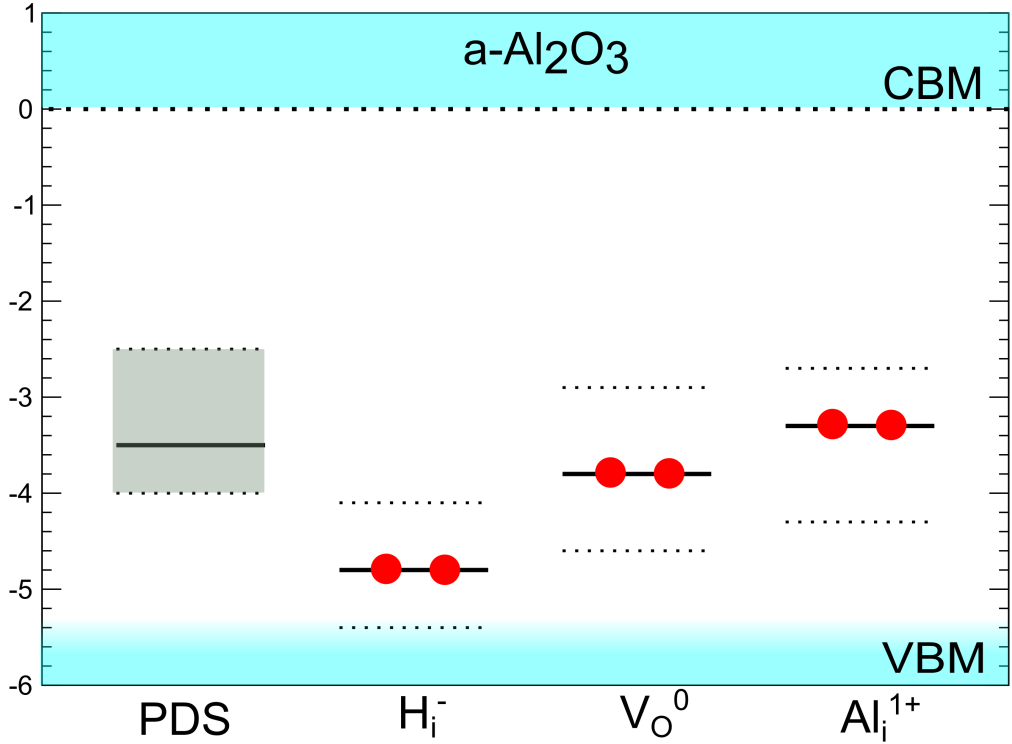}
\caption{\label{fig:KSlevels} The average (solid black line) and range (dotted lines) of the K-S energy levels of the Al$_{\textrm{i}}$, V$_{\textrm{O}}$ or H$_{\textrm{i}}$ defects with respect to the a-Al$_2$O$_3$ conduction band, compared to the experimental EPDS trap energy level range~\cite{Zahid2010}. For the EPDS data the solid black line shows the energy of  the maximum density of states seen in Fig. \ref{fig:PDS}.}
\end{figure}

\subsection{Oxygen vacancies}

\subsubsection{V$_{\textrm{O}}$ in $\alpha$-Al$_2$O$_3$}

There is a large body of existing literature on oxygen vacancies in crystalline Al$_2$O$_3$, both computational~\cite{Liu2010,Weber2011,Choi2013a} and experimental~\cite{Springis1984,Pustovarov2010}. This allows calculations of  V$_{\textrm{O}}$ in $\alpha$-Al$_2$O$_3$ to be used  to benchmark the DFT setup and hybrid functional parameters with respect to existing studies, and to act as a point of comparison to the amorphous system.

\begin{figure}
\centering
\includegraphics[width=1.0\linewidth]{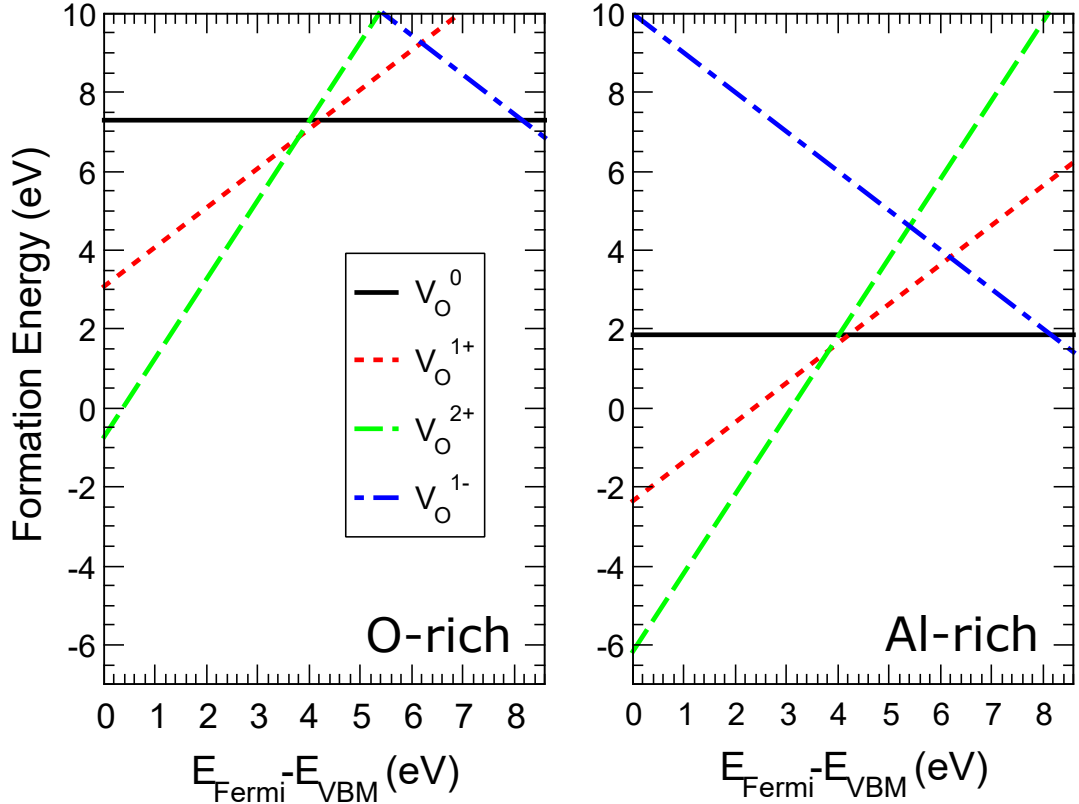}
\caption{\label{fig:alphaVoEform} The formation energy of the different charge states of oxygen vacancies in $\alpha$-Al$_2$O$_3$ as a function of the Fermi energy position with respect to the valence band.}
\label{single}
\end{figure}

Fig. \ref{fig:alphaVoEform} shows the formation energy of various charge states of oxygen vacancies in $\alpha$-Al$_2$O$_3$ calculated in this work. one can see that oxygen vacancies have a (+2/+1) charge transition level at 3.8 eV above the VBM, with the (+1/0) level at 4.0 eV above the VBM, meaning that for most Fermi energies in the band gap the V$_{\textrm{O}}^{2+}$ and V$_{\textrm{O}}^{0}$ are the most thermodynamically stable charge states of the oxygen vacancy. During the geometry relaxation of the neutrally charged oxygen vacancy, the four nearest neighbor Al move 0.01-0.10 $\textrm{\AA}$ towards the defect site with respect to the perfect lattice. For  V$_{\textrm{O}}^{2+}$ they relax 0.19-0.30 $\textrm{\AA}$ away from the vacancy site with respect to the perfect lattice. These displacements are in good agreement with Choi et al.~\cite{Choi2013a}. However, whilst the nearest neighbor displacements are similar, they calculate the (+2/+1) charge transition level to be 3.2 eV above the VBM, with the (+1/0) level at 4.1 eV~\cite{Choi2013a}. This difference is likely to stem from the smaller cell size used in~\cite{Choi2013a} (160 atoms), which constrains the relaxation of the next nearest neighbor ions, increasing the energy of the V$_{\textrm{O}}^{2+}$ defect. The over-estimated V$_{\textrm{O}}^{2+}$ formation energy has also been obtained in other calculations using smaller cell sizes~\cite{Liu2010}.

\begin{figure}
\centering
\includegraphics[width=1.0\linewidth]{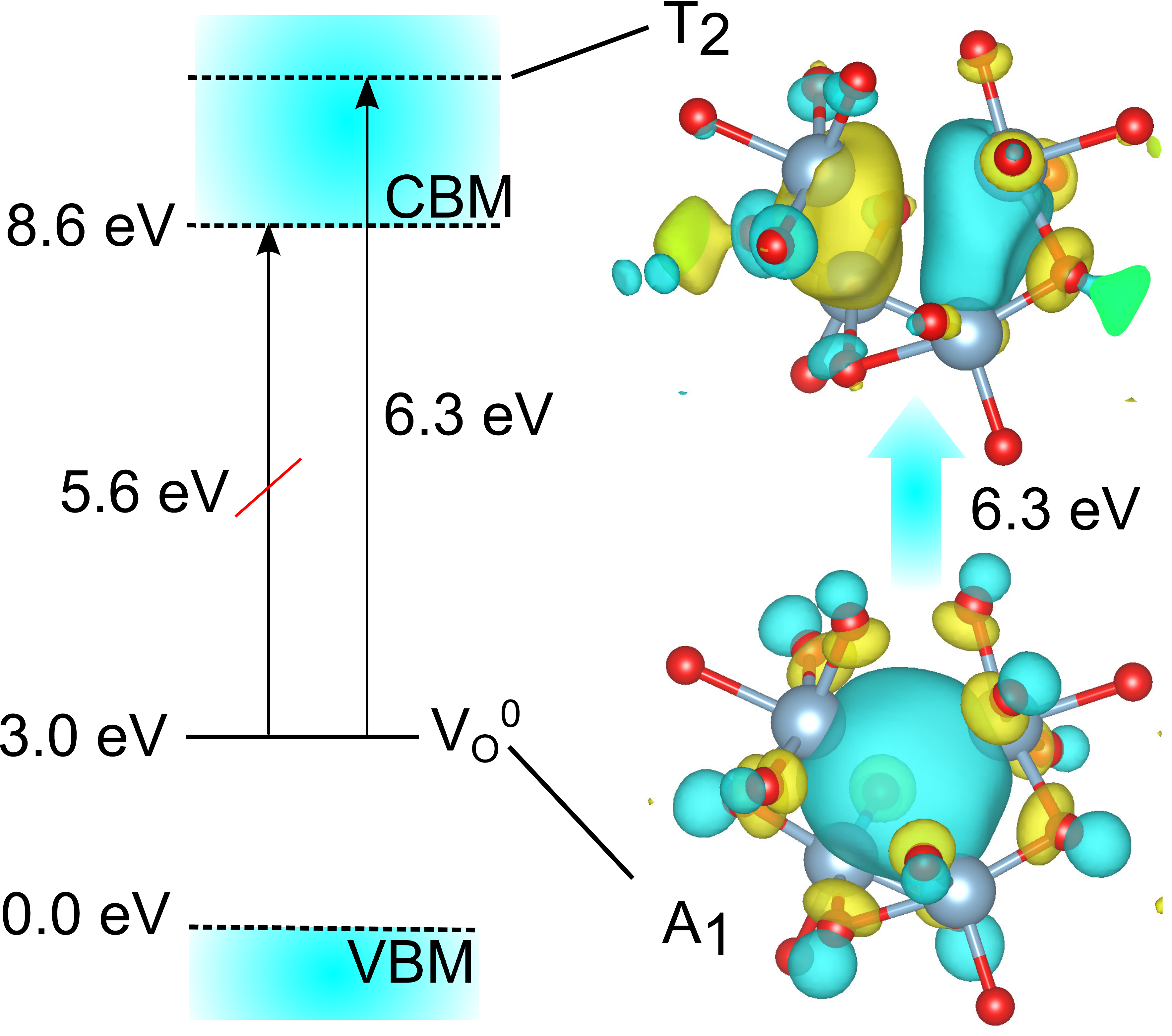}
\caption{\label{fig:alphaVotrans} An energy level diagram for V$_{\textrm{O}}^{0}$ in $\alpha$-Al$_2$O$_3$ showing the symmetry forbidden and allowed transitions, and the molecular orbitals of the A$_1$ and T$_2$ states involved in the symmetry allowed excitations.}
\label{single}
\end{figure}

Spectroscopic measurements~\cite{Pustovarov2010} of $\alpha$-Al$_2$O$_3$ assign an absorption peak at 6.4 eV to the neutral oxygen vacancy. This energy is greater than the 5.6 eV energy difference calculated between the V$_{\textrm{O}}^{0}$ K-S energy level and the conduction band minimum. However, O ions in $\alpha$-Al$_2$O$_3$ are four-coordinated and have tetrahedral like symmetry with point group T$_{\textrm{d}}$. The ground state of V$_{\textrm{O}}^{0}$ has the A$_1$ like character (see Fig. \ref{fig:alphaVotrans}) which has minimal wavefunction overlap with the delocalized CBM in $\alpha$-Al$_2$O$_3$ composed of Al 3$s$ orbitals. However, as can be seen in Fig. \ref{fig:alphaVotrans}, the defect induces a localized state in the conduction band with T$_2$ like character. A$_1$ to T$_2$ excitations are symmetry allowed dipole transitions and are most likely responsible for the sharp peak observed experimentally~\cite{Pustovarov2010}. The T$_2$ like state (see Fig. \ref{fig:alphaVotrans}) lies 6.3 eV above the V$_{\textrm{O}}^{0}$ ground state K-S level, and, although the energy difference between K-S levels is only a first approximation for excitation energies, the calculated transition energy is in a very good agreement with the experimentally observed peak position attributed to V$_{\textrm{O}}^{0}$. 

\subsubsection{V$_{\textrm{O}}$ in a-Al$_2$O$_3$}

\begin{figure}
\centering
\includegraphics[width=1.0\linewidth]{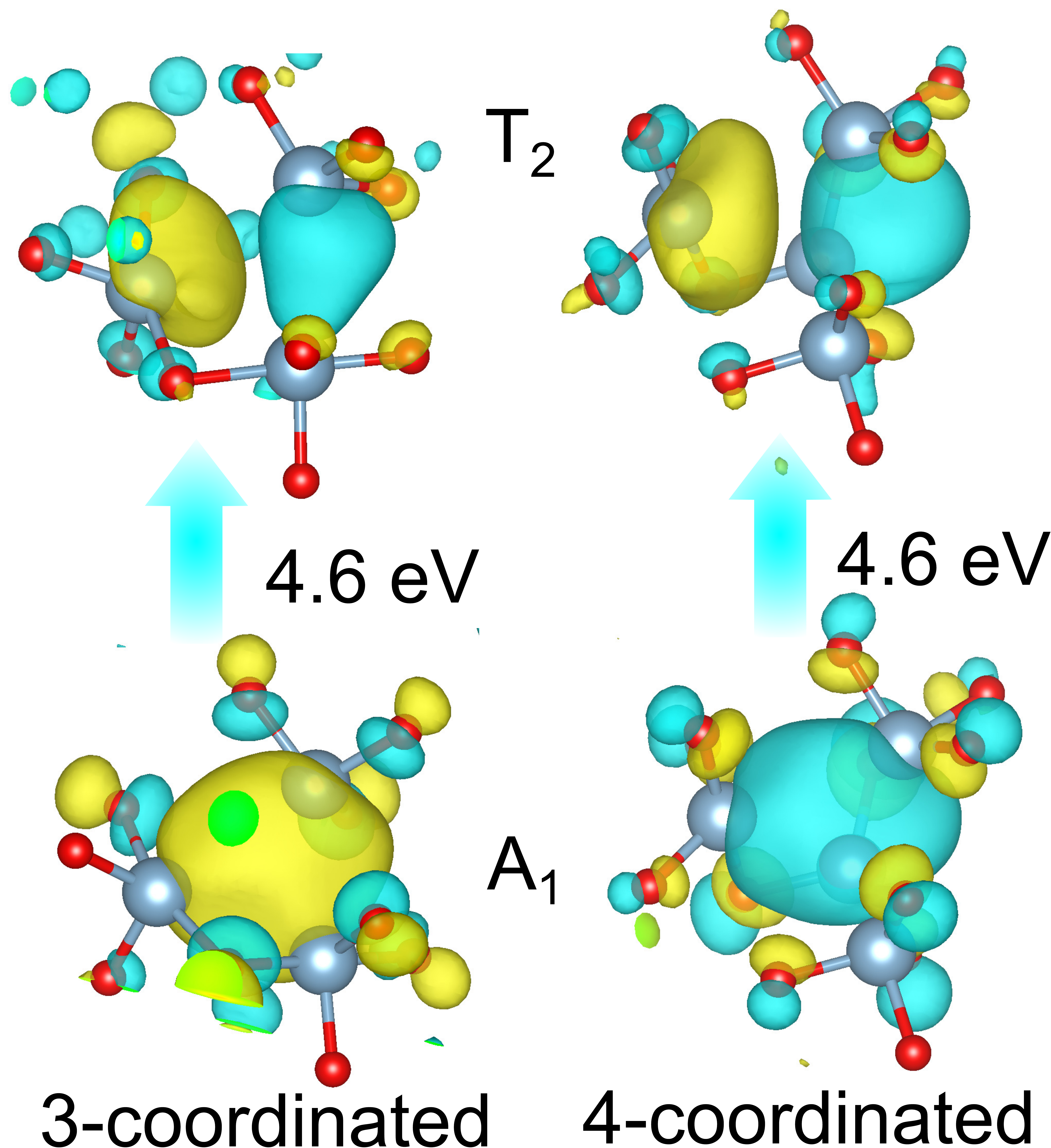}
\caption{\label{fig:aVotrans} The molecular orbitals of the A$_1$ and T$_2$ like states of 3 and 4 coordinated V$_{\textrm{O}}^{0}$ in a-Al$_2$O$_3$, and difference in energy between their Kohn-Sham levels.}
\label{single}
\end{figure}

Oxygen vacancies in amorphous Al$_2$O$_3$ have also been investigated using DFT methods~\cite{Liu2013,Guo2016}. However, these studies only model vacancies in a single cell of 80~\cite{Liu2013} and 160~\cite{Guo2016} atoms respectively, which may not capture the full range of properties of V$_{\textrm{O}}$. In order to improve understanding of these defects in a-Al$_2$O$_3$, the properties of oxygen vacancies at 11 defect sites in 10 geometry samples have been calculated and are presented here. The distribution of the O coordination numbers has been taken into consideration, with 7 $^{[3]}$O, 2 $^{[2]}$O and 2 $^{[4]}$O being removed in order to create the oxygen vacancies.

In amorphous Al$_2$O$_3$ only the +2 or neutral charge states of the oxygen vacancy are thermodynamically stable, as also observed by Guo et al.~\cite{Guo2016}. The (+2/0) charge transition level lies, on average, 3.5 eV above the VBM and 2.0 eV below the CBM (see Fig. \ref{fig:qtrans}). In the neutral charge state 2 electrons localize on the vacancy site, forming an F-center, similar to the defect in the crystalline system. As V$_{\textrm{O}}^{0}$ has a doubly occupied state in the band gap it could be responsible for the transitions seen experimentally~\cite{Zahid2010}, after charge injection. The high (+2/0) charge transition level suggests that before charge injection, whilst the Fermi level lies at the VBM, V$_{\textrm{O}}^{2+}$ is the most stable configuration, which has no occupied states in the band gap. In a-Al$_2$O$_3$ there is no stable V$_{\textrm{O}}^{1-}$ state and so it cannot be responsible for the negative charging observed in amorphous alumina films, unless it acts to compensate the negative charge of another defect.

The calculated  K-S energy levels of V$_{\textrm{O}}^{0}$ lie on average at 4.0 eV below the CBM (see Fig. \ref{fig:KSlevels}). The defects in the amorphous material create similar states to those observed in $\alpha$-Al$_2$O$_3$. The 3- and 4-coordinated vacancies induce localized  states, similar to the A$_1$ state, to form in the band gap, and T$_2$ like states to appear within the conduction band (see Fig. \ref{fig:aVotrans}). The further distortion of the tetrahedral symmetry means there is greater mixing of the states, and at 3-coordinated sites, there is significant relaxation from the next nearest neighbor Al ions, but the presence of the T$_2$ like state suggests there could be high oscillator strength transitions at larger excitation energies. On average the K-S energy levels of the T$_2$ like states lie 4.8 eV above the doubly-occupied energy level of V$_{\textrm{O}}^{0}$. This energy difference is significantly lower than the 6.5 eV excitation energy assigned to the neutral oxygen vacancy measured using electron energy loss spectroscopy (EELS)~\cite{Perevalov2010}. However, the degree of crystallinity of the measured samples is unclear and in partially crystalline samples this peak could correspond to the onset of inter-band transitions or excitation of V$_{\textrm{O}}^{0}$ states in crystalline Al$_2$O$_3$~\cite{Perevalov2010}.

\subsection{Interstitial oxygen}

\subsubsection{O$_i$ in $\alpha$-Al$_2$O$_3$}

\begin{figure}
\centering
\includegraphics[width=1.0\linewidth]{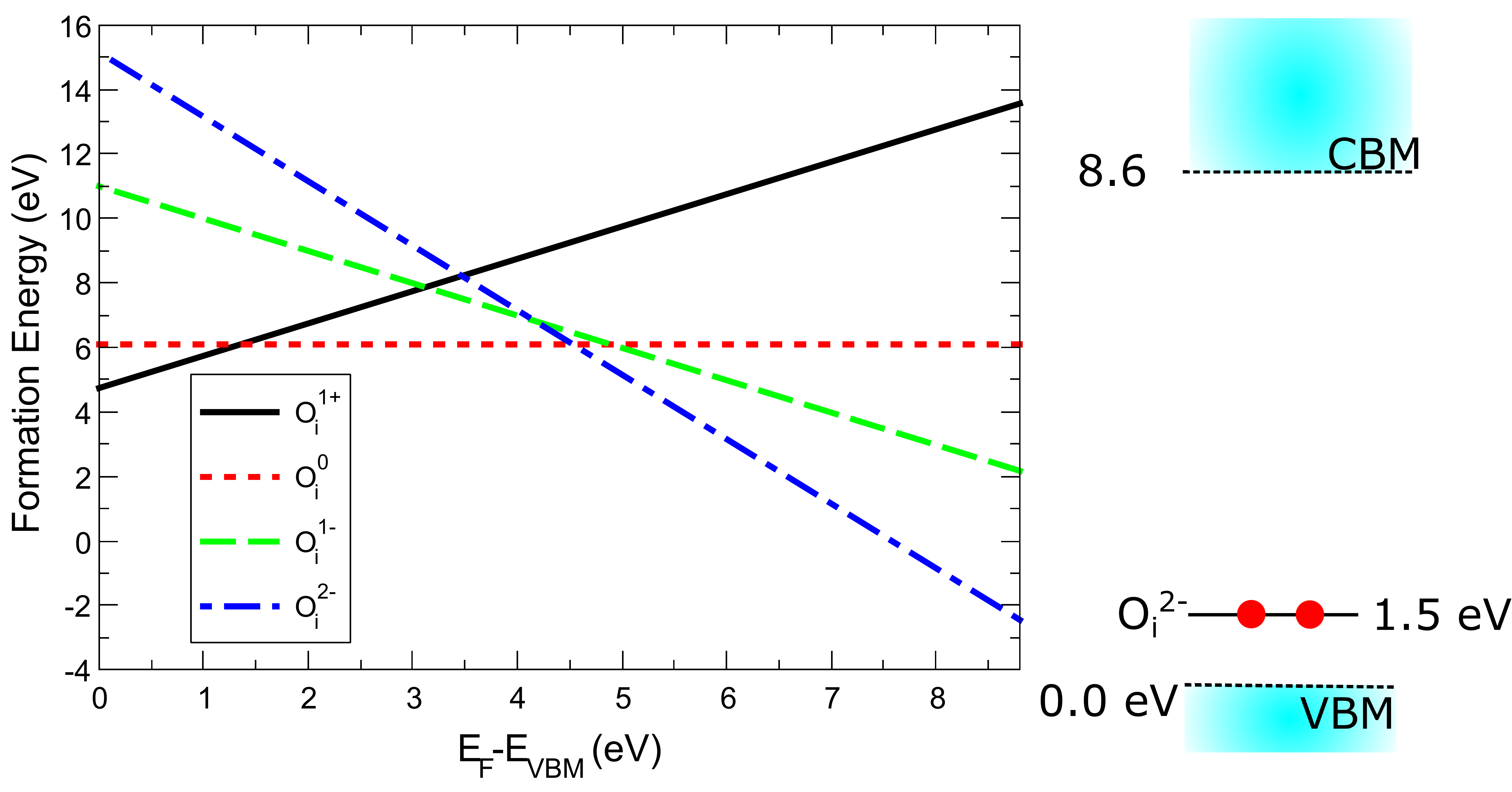}
\caption{\label{fig:alphaOi} The formation energy of the different charge states of interstitial oxygen in $\alpha$-Al$_2$O$_3$ against the Fermi energy with respect to the valence band, calculated in the O-rich condition. On the right is the Kohn-Sham energy level position of the doubly occupied O$_{\textrm{i}}^{2-}$ defect. }
\label{single}
\end{figure}

Neutrally charged oxygen interstitials in $\alpha$-Al$_2$O$_3$ form an O-O dimer at the defect site, centred on the original lattice position of the O ion. In this configuration both oxygens are 3 coordinated with Al, and the O-O bond length is 1.44 $\textrm{\AA}$. It has 3 thermodynamically stable charge states, as can be seen from the formation energy diagram in Fig. \ref{fig:alphaOi}. Importantly, it traps electrons and becomes O$_{\textrm{i}}^{2-}$ when the Fermi energy is 4.5 eV above the VBM. This is close to the 4.7 eV (0/-2) charge transition level calculated by Choi et al.~\cite{Choi2013a}, using the HSE06 functional. 

When the oxygen interstitial captures 2 electrons to become O$_{\textrm{i}}^{2-}$, the O-O bond length increases to 2.15 $\textrm{\AA}$. This large displacement results in both O ions becoming 4 coordinated with Al, forming tetrahedral configurations. Due to this relaxation the K-S energy level of O$_{\textrm{i}}^{2-}$ lies only 1.5 eV above the valence band (see Fig. \ref{fig:alphaOi}), as both O are fully coordinated. Its role as a deep acceptor is in good agreement with previous calculations~\cite{Fonseca2008}.

\subsubsection{O$_{\textrm{i}}$ in a-Al$_2$O$_3$}\label{aOi}

To model the oxygen interstitial in the amorphous structure, single oxygen atoms were added to the 10 different geometry samples and placed within 1.6 $\textrm{\AA}$ of an O ion, with 4 near $^{[3]}$O, 3 near $^{[2]}$O and 3 next to $^{[4]}$O. Similar to the crystalline case, oxygen interstitials in a-Al$_2$O$_3$ have a deep (0/-2) charge transfer level. The average charge transfer energy lies 3.5 eV below the CBM, and 2.0 eV above the VBM. Guo et al.~\cite{Guo2016} calculated the average charge transfer level to be 2.5 eV above the VBM and similarly show that the O$_{\textrm{i}}^{-}$ is thermodynamically unstable.

The fully relaxed neutral oxygen interstitial forms an O-O peroxy bond with the nearest neighbor oxygen. The average O-O bond length of the 10 defect configurations is 1.46 $\textrm{\AA}$ for the neutral charge state. When 2 electrons are added to the system, forming O$_{\textrm{i}}^{2-}$, the O-O bond elongates to an average of 2.40 $\textrm{\AA}$. This large relaxation significantly lowers the energy of the defect-induced $\sigma_{2p}^{*}$ like orbitals in the conduction band, down towards the valence band. All the occupied K-S energy levels of the O$_{\textrm{i}}^{2-}$ lie within the valence band, with no states existing in the band gap. The low lying charge transition level of interstitial oxygen means it is likely to be a source of electron trapping in a-Al$_2$O$_3$, but, the lack of states in the band gap means that it cannot explain the trap spectroscopy data~\cite{Zahid2010}. 
\subsection{Aluminium vacancies}

\subsubsection{V$_{\textrm{Al}}$ in $\alpha$-Al$_2$O$_3$}

\begin{figure}
\centering
\includegraphics[width=1.0\linewidth]{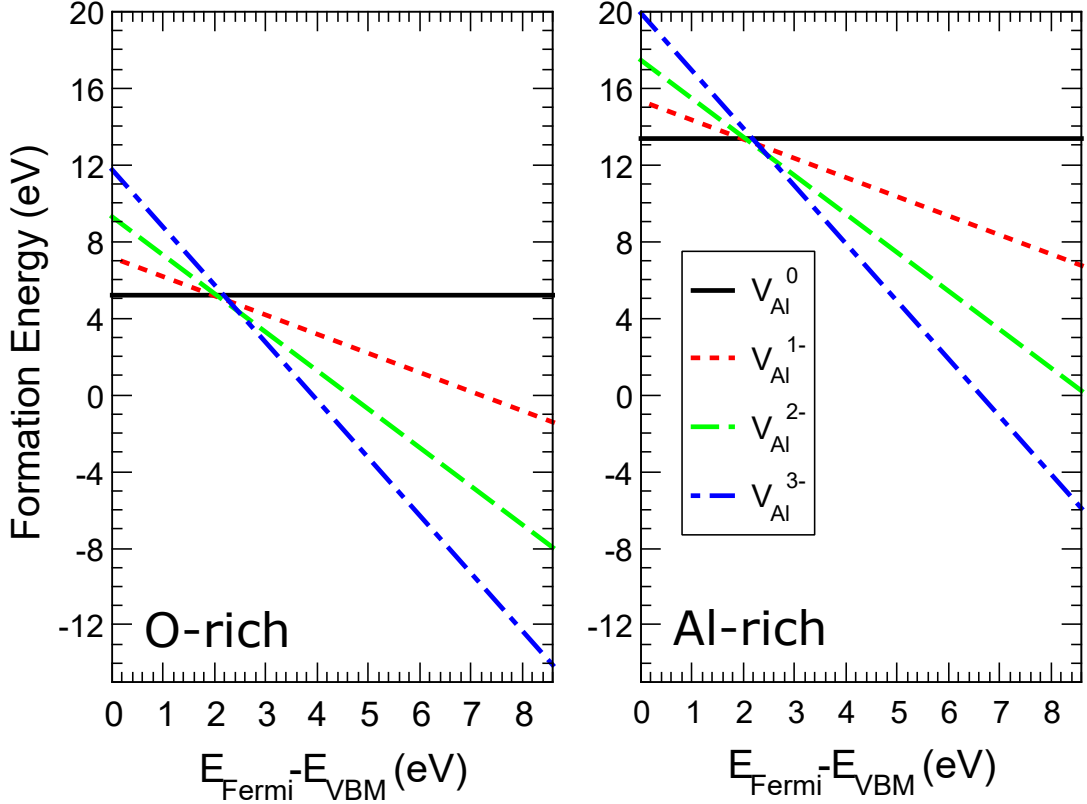}
\caption{\label{fig:alphaVAl} The formation energy of the different charge states of aluminium vacancies in $\alpha$-Al$_2$O$_3$ plotted against the Fermi energy with respect to the valence band.}
\label{single}
\end{figure}

The formation energy diagram for V$_{\textrm{Al}}$ in $\alpha$-Al$_2$O$_3$ (see Fig. \ref{fig:alphaVAl}) shows that the -3 charge state of the vacancy becomes stable at 2.4 eV above the VBM. This means the Al vacancy (0/-3) charge transition level lies even lower in the band gap than the interstitial oxygen (0/-2) transition level (see Fig. \ref{fig:alphaOi}), which suggests that it will be the dominant negatively charged defect in crystalline alumina, even in O-rich conditions. This implies that it is therefore a likely source of fixed negative charge in amorphous alumina. This agrees with previous studies that show Al vacancies in $\alpha$-Al$_2$O$_3$~\cite{Choi2013a} and $\kappa$-Al$_2$O$_3$~\cite{Choi2013a,Weber2011} are very deep acceptors, with charge transition levels close to the valence band.

\subsubsection{V$_{\textrm{Al}}$ in a-Al$_2$O$_3$}

As can be seen in Fig \ref{fig:qtrans}, the -3 charge state of Al vacancies in a-Al$_2$O$_3$ becomes stable when Fermi energies are on average 3.5 eV below the conduction band (2.0 eV above the valence band). This is the lowest lying charge transition level of all the defects presented in this paper, but it is very close to the O$_{\textrm{i}}$ (0/-2) level. 10 vacancy sites were examined with 4 $^{[4]}$Al, 3 $^{[4]}$Al and 3 $^{[4]}$Al removed to create the defects. However, little dependence on coordination was observed, with a deviation in the average (-2/-3) charge transfer level of less than 0.5 eV.

It is likely that Al vacancies will acts as deep electron traps in a-Al$_2$O$_3$, but, the highest occupied K-S energy level of V$_{\textrm{Al}}^{3-}$ (across all the samples) lies 4.7 eV below the CBM, with most of the defect states lying within the valence band. This suggests that it is unlikely to be the charge trap measured by Zahid et al.~\cite{Zahid2010}. It is more likely to act as a source of negative charge that compensates for positively charged defects before electron injection, similar to the mechanism described in section \ref{aOi}.

\subsection{Interstitial Al}

\subsubsection{Al$_{\textrm{i}}$ in $\alpha$-Al$_2$O$_3$}

\begin{figure}
\centering
\includegraphics[width=1.0\linewidth]{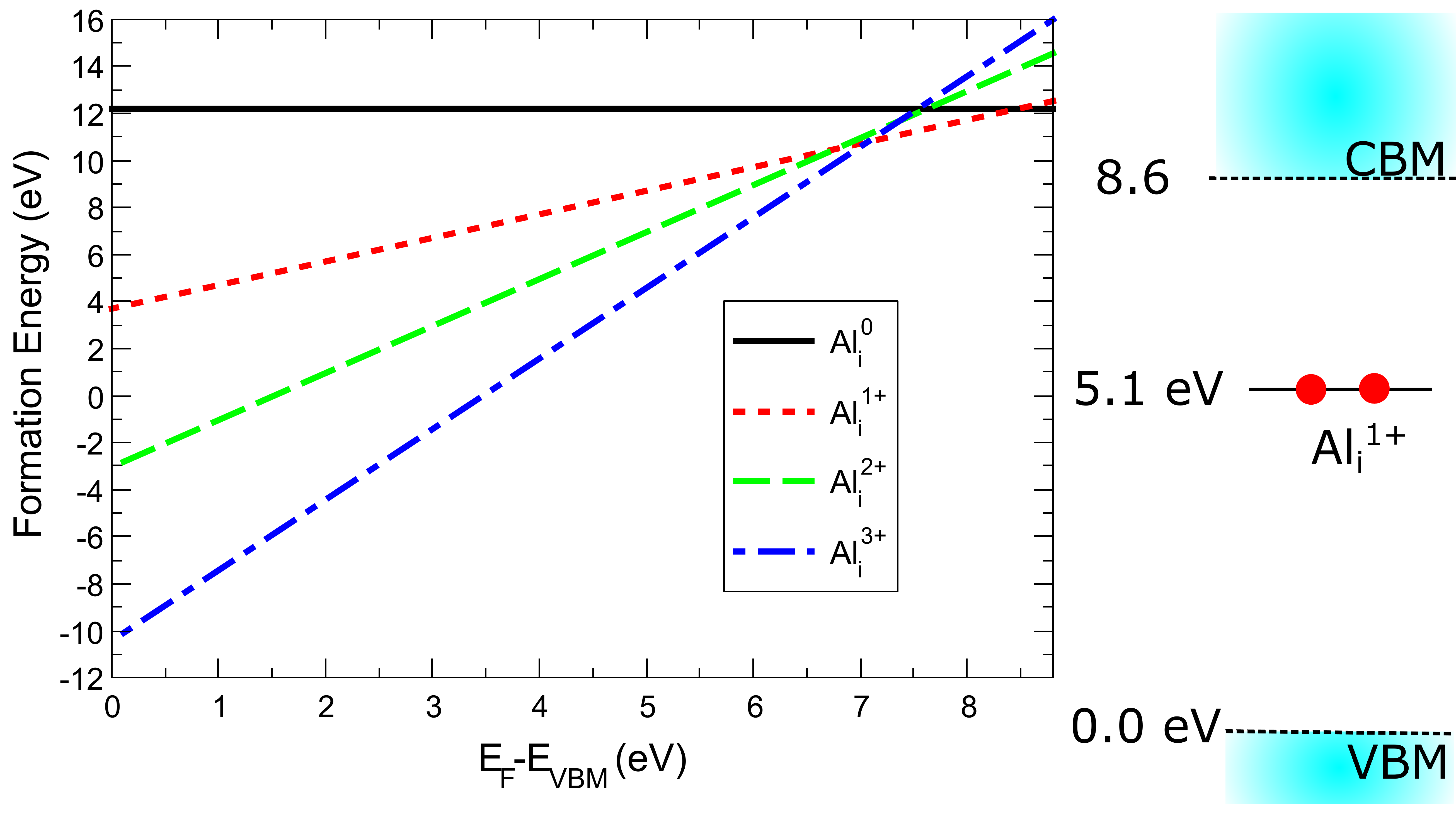}
\caption{\label{fig:alphaAli} The formation energy of interstitial aluminium and its various charge states in $\alpha$-Al$_2$O$_3$, plotted against the Fermi energy with respect to the valence band, calculated in the Al-rich environment.}
\label{single}
\end{figure}

On a first examination of Al interstitials in $\alpha$-Al$_2$O$_3$, they do not appear to be a good candidate for the negative charging observed experimentally. The formal charge of Al in Al$_2$O$_3$ is 3+, and Bader analysis shows the system is highly ionic. Thus, the addition of an Al atom donates 3 electrons into the system without introducing any unoccupied states in the predominantly O 2$p$ valence band, meaning the defect is most likely to act as a donor in $\alpha$-Al$_2$O$_3$. This is demonstrated by the formation energy diagram shown in Fig. \ref{fig:alphaAli} where the (+3/+1) level is only 1.6 eV below the CBM. Al$_{\textrm{i}}^{3+}$ has the lowest formation energy for a wide range of the Fermi energy, and no occupied states in the band gap available for excitations into the conduction band. There are no thermodynamically stable negative charge states of Al$_{\textrm{i}}$ observed at any Fermi energy, to act as independent electron traps.

\begin{figure}
\centering
\includegraphics[width=1.0\linewidth]{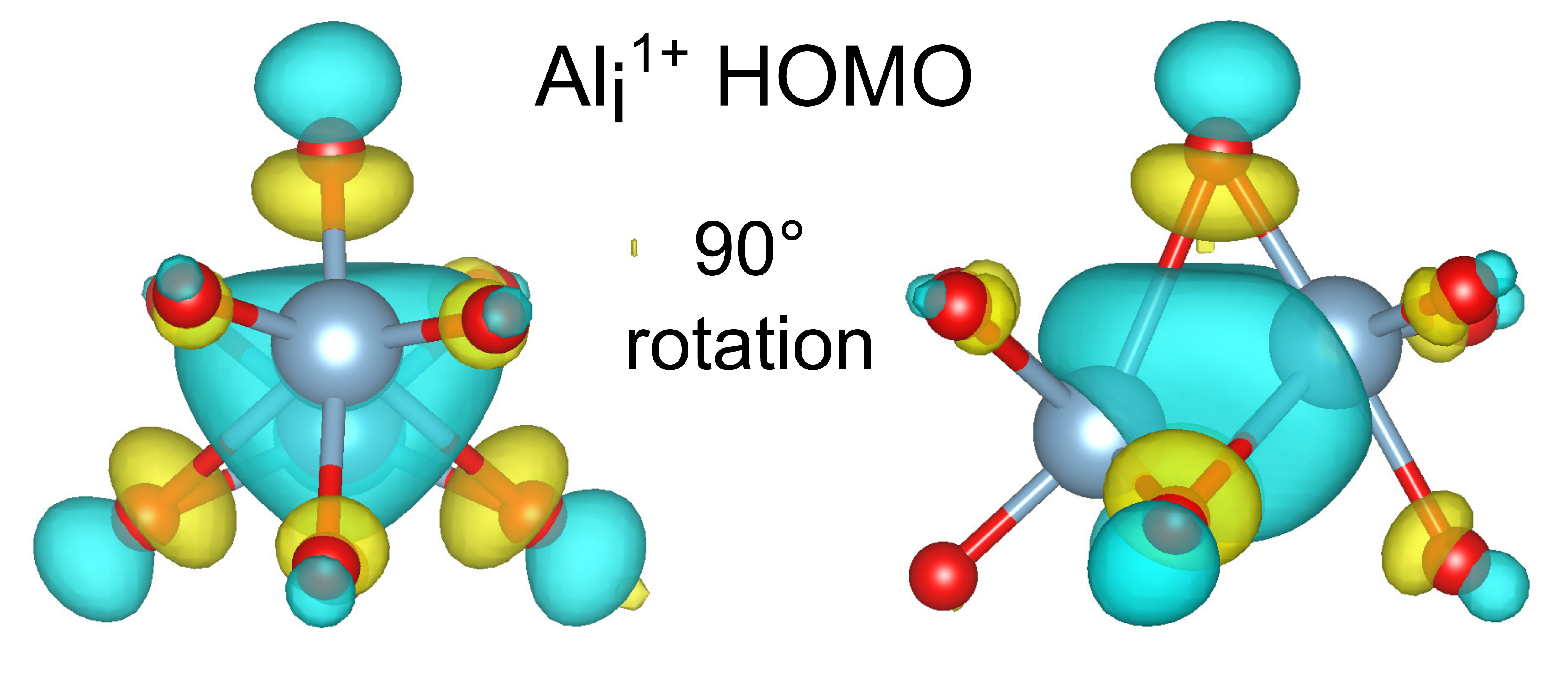}
\caption{\label{fig:alphaAliq1HOMO} The iso-surface of the Al$_{\textrm{i}}^{1+}$ HOMO, in 2 orientations so as to see the C$_{3h}$ point symmetry. The interstitial Al and its nearest neighbor Al are both 6 coordinated with O, though some Al-O bonds are extended to 2.6 $\textrm{\AA}$.}
\label{single}
\end{figure}

However, Al$_{\textrm{i}}^{1+}$ has a doubly occupied KS energy level in the middle of the band gap, 3.5 eV below the CBM (see Fig. \ref{fig:alphaVAl}). The C$_{3h}$ like point symmetry of the Al$_{\textrm{i}}^{1+}$ highest occupied molecular orbital (HOMO), can be seen in Fig. \ref{fig:alphaAliq1HOMO}. The symmetry is a result of the triangle of oxygen ions whose atomic orbitals point into the defect center between the 2 Al ions. The electrons localize between the 2 positively charged Al ions which lowers its energy. This symmetry means that an excitation into the conduction band minimum is a dipole allowed transition (A' to E'). Thus, at least for the crystalline system, there exists a mid gap state in the same energy range as the levels seen experimentally~\cite{Zahid2010}, and, unlike the oxygen vacancy, the defect state perturbs the CBM state and an occupying electron can be excited straight into the bottom of the conduction band.

\subsubsection{Al$_{\textrm{i}}$ in a-Al$_2$O$_3$}

In a-Al$_2$O$_3$ the average KS energy level of Al$_{\textrm{i}}^{1+}$ lies 3.4 eV below the conduction band, with a range of 2.7-4.3 eV  (see Fig. \ref{fig:KSlevels}), in very good agreement with both the EPDS and GS-TSCIS measurements~\cite{Zahid2010}. The average (+3/+1) charge transition level is 2.1 eV below the CBM (see Fig. \ref{fig:qtrans}). This means that for a large range of the Fermi energy interstitial Al is most stable in the 3+ charge state, which has no occupied states in the gap. However, after electron injection, the unoccupied states of the Al$_{\textrm{i}}^{3+}$ will trap electrons and become Al$_{\textrm{i}}^{1+}$. Al$_{\textrm{i}}^{1+}$ has a doubly occupied state in the gap, meaning the now filled gap states can then be excited into the conduction band and electrons detected. 

\subsection{Comparison with the spectroscopic data}

Our results suggest that O$_{\textrm{i}}$ and V$_{\textrm{Al}}$ can be responsible for negative charging of a-Al$_2$O$_3$ films. Both defects become negatively charged when the Fermi energy is approximately in the middle of the band gap, and, as can be seen in Fig. \ref{fig:qtrans}, their average charge transition levels lie below the Si VBM. This means that they are likely to be in the negative charge state for most values of the Fermi energy. However, they do not have occupied levels in the band gap which could be depopulated in EPDS measurements~\cite{Zahid2010}. Therefore, although they carry negative charge, they are most likely not responsible for the trap states observed in ref.~\cite{Zahid2010}. On the other hand,  Al$_{\textrm{i}}^{1+}$ and V$_{\textrm{O}}^{0}$ centers have occupied states in the gap with energies in the range observed in ref.~\cite{Zahid2010}. These results, as well as the large width of the EPDS spectrum in  Fig. \ref{fig:PDS}, suggest that a combination of different types of compensating defects, rather than a single defect, is more likely to be responsible for the behavior seen experimentally and the EPDS data. 

For example, one can assume that in low density a-Al$_2$O$_3$ some Al$_2$O$_3$ molecular units are missing, implying the existence of compensating O and Al vacancies V$_{\textrm{O}}^{2+}$ and V$_{\textrm{Al}}^{3-}$. As discussed above, V$_{\textrm{Al}}^{3-}$ are unlikely to be observed by EPDS as they do not have occupied states in the right energy range. However, after trapping two extra electrons at the V$_{\textrm{O}}^{2+}$ site the whole system becomes negatively charged, with V$_{\textrm{O}}^{0}$ having doubly occupied levels at the energies corresponding to the EPDS spectra in  Fig. \ref{fig:PDS}.

The much lower density of negatively charged states observed in the gap before the electron injection~\cite{Zahid2010} (see Fig. \ref{fig:PDS}) can be explained by an electron tunneling from the Si substrate and charging of the interface layer. Negatively charged O$_{\textrm{i}}$ or V$_{\textrm{Al}}$ are then only partially compensated by positively charged Al$_{\textrm{i}}$, V$_{\textrm{O}}$ or H$_{\textrm{i}}$ defects, making the system slightly negative with a small density of states in the gap. The electron injection occurs at high positive voltage by electron tunneling through the SiO$_2$ tunneling
layer. This means that the system's Fermi level is higher than the Si/a-Al$_2$O$_3$ band offset of 2.1 eV below the bottom of the a-Al$_2$O$_3$ conduction band \cite{Afanas'ev_2007_IPErev}. Therefore after the electron injection, more electrons occupy the Al$_{\textrm{i}}$, V$_{\textrm{O}}$ or H$_{\textrm{i}}$ states in the gap, which are then observed spectroscopically (see Fig. \ref{fig:PDS}) via excitation into the conduction band. This filling of states also leads to a much larger overall negative charge being observed. This hypothesis is supported by measurements of electron capture cross sections that suggest that electrons trap at positively charged defect centers~\cite{Afanasev2004} in a-Al$_2$O$_3$, and that both positively and negatively charged defect centers are simultaneously present in the material. We also note that after electron injection all defect states are diamagnetic, which explains the absence of ESR signatures discussed in the Introduction.

\begin{figure}
\centering
\includegraphics[width=1.0\linewidth]{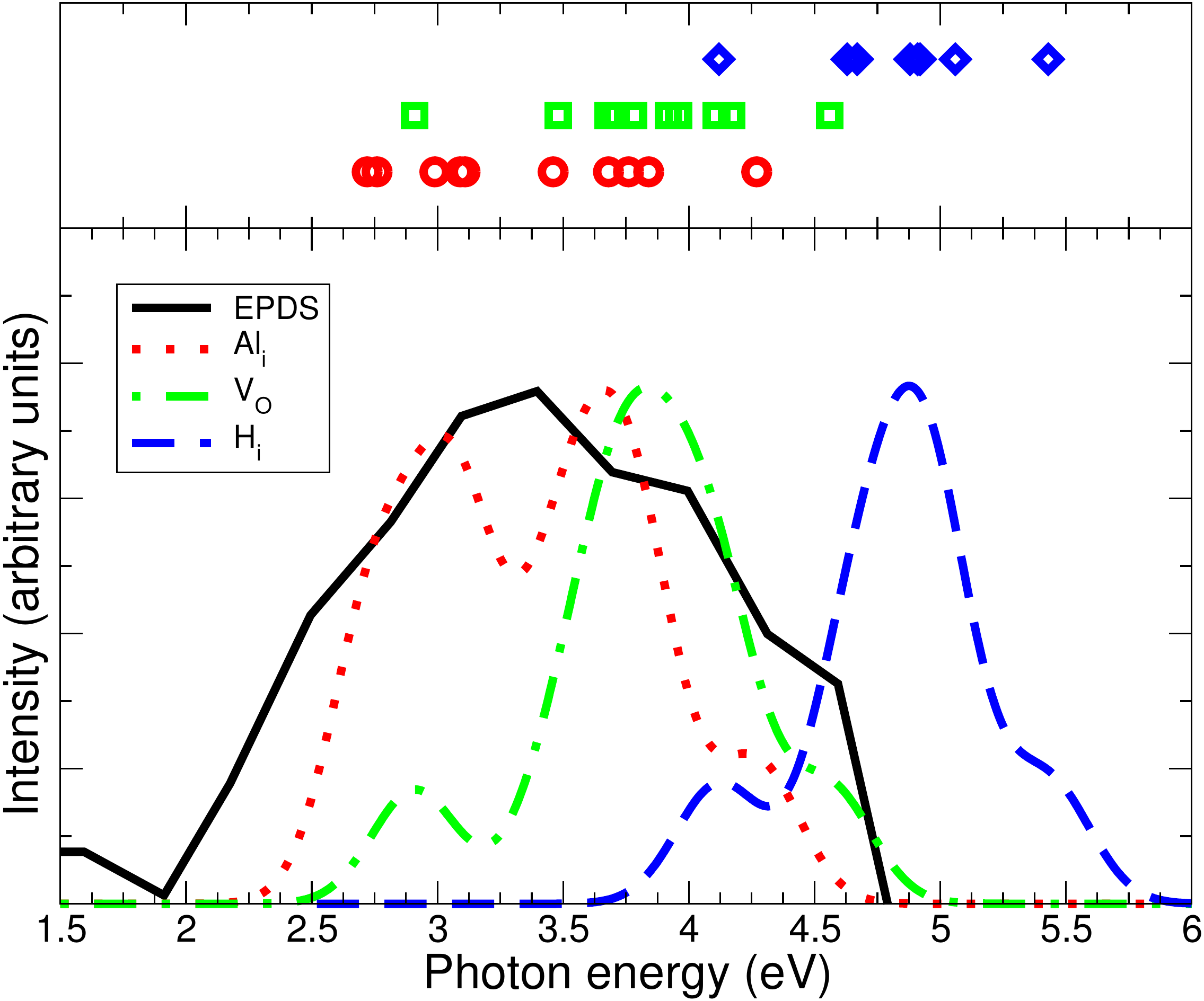}
\caption{\label{fig:totalspectra} The transition energy distribution of the Al$_{\textrm{i}}$ (circle), V$_{\textrm{O}}$ (square) and H$_{\textrm{i}}$ (diamond) defects, predicted using the differences in energies between the Kohn-Sham energy levels of the in-gap occupied states and the conduction band state, as seen in Fig. \ref{fig:KSlevels}. The calculated transition energy of each defect from all 10 cells is also broadened using a Gaussian with a full width half maxima of 0.4 eV, normalized and compared to the EPDS data~\cite{Zahid2010} (black line in figure) shown in Fig. \ref{fig:PDS}, in order to give an indication of the grouping of the levels. }
\label{single}
\end{figure}

To further compare the calculated energies with the EPDS spectra~\cite{Zahid2010,afanas2014invited}, as a first approximation of the photo-excitation energy, we use the energy difference between the K-S energy levels of the defects and the CBM calculated using DFT. Time dependent DFT (TDDFT) calculations have demonstrated that the excitation energies of transitions from localized defect states into delocalized band states can be well approximated by the K-S energy  differences~\cite{BERNASCONI2004141,Strand2017b}. This is attributed to the exact exchange related electron-hole interaction vanishing when one of the states is delocalized, meaning the transition energy is mainly determined by the K-S energies~\cite{BERNASCONI2004141}. In ~\cite{Dicks2017} it is shown that the CBM of a-Al$_2$O$_3$ is a delocalized state, and does not exhibit the electron localization seen in other oxides~\cite{Kaviani2016}. High electron mobilities in a-Al$_2$O$_3$ have also been measured experimentally~\cite{Novikov2009}. As EPDS detects transitions by measuring charge loss due to electron drift in the alumina CB, it is important that the energy reference is taken from the mobility edge, which in a-Al$_2$O$_3$ corresponds to the CBM~\cite{Novikov2009}. With this in mind, the range of K-S energy levels of Al$_{\textrm{i}}$, V$_{\textrm{O}}$ or H$_{\textrm{i}}$, with respect to the CBM, shown in Fig. \ref{fig:KSlevels}, all have levels within the range measured experimentally~\cite{Zahid2010}. The energy level distribution of these K-S levels with respect to the conduction band can be seen in Fig. \ref{fig:totalspectra}, where each predicted transition energy, for each defect, has been broadened using a Gaussian with a full width half maxima of 0.4 eV and normalized to all have the same maximum. The distribution of energies of the Al$_{\textrm{i}}$ and V$_{\textrm{O}}$ in Fig. \ref{fig:totalspectra} match those of the EPDS measurements~\cite{Zahid2010} well, but the distribution of H$_{\textrm{i}}$ defect levels mostly falls outside the measured spectra and so is less likely to contribute to the transitions observed. This means more than one type of 'intrinsic' defect could be responsible for the trap states, but hydrogen implantation is less likely to affect the EPDS measurements. Experimentally the defect species could be more confidently assigned by adjusting the growth conditions of alumina thin films so as to control the O and Al chemical potentials.

In this study we specifically addressed the negative charging of alumina since experiments clearly show that this is the dominant process responsible for the violation of electroneutrality in this dielectric. It is worth noting that even in the case of hole injection in alumina thin films which have been deposited on oxidized silicon, the dominant positive charging is associated with the SiO$_2$ interlayer, rather than the alumina layer~\cite{Afanasev2004}. Similar conclusions regarding SiO$_2$ dominated positive charging, correlated with protonic contribution, have also been reached in the past for atomic-layer deposited ZrO$_2$~\cite{Afanasev2004,Houssa2000,Houssa2001} and HfO$_2$~\cite{Afanasev2004} insulators. 
Intrinsic positive charging has been reported in HfO$_2$ when the contribution of the SiO$_2$ layer is excluded by depositing the metal oxide on top of other materials~\cite{Jurczak2011}. This correlates with the development of a significant oxygen deficiency due to application of an O-scavenging metal (Hf) layer~\cite{Jurczak2011}.
This observation is consistent with the results of our previous calculations indicating that the oxygen vacancy in hafnia favours the positively charged state when the electron chemical potential is lower than approximately 1.5 eV below the conduction band minimum~\cite{Cerbu2016}. Therefore, we believe that oxygen deficiency in metal oxide insulators represents the major factor enabling the formation of positive charge. Since the O vacancy formation in Al$_2$O$_3$ is less energetically favorable than in other high-permittivity metal oxides (ZrO$_2$, HfO$_2$, SrTiO$_3$, etc.), contribution of this positive charge to the net oxide charge observed experimentally will be less pronounced leading to the dominance of negative charging. Nevertheless, experiments on the internal photoemission of electrons into $\gamma$-Al$_2$,O$_3$ layers in which oxygen deficiency is developed via the deposition of Ti (which is also an oxygen scavenger) suggest charging of the interface~\cite{Filatova2016} similar to that observed in HfO$_2$ upon O scavenging by metallic Hf~\cite{Jurczak2011}.

\section{Conclusion}

In order to understand the source of negative charging in a-Al$_2$O$_3$ films~\cite{Govoreanu2006,Novikov2009,Zahid2010,Li2014}, the electronic properties of the defects H$_{\textrm{i}}$, V$_{\textrm{O}}$, O$_{\textrm{i}}$, V$_{\textrm{Al}}$ and Al$_{\textrm{i}}$ were calculated using DFT. 

O$_{\textrm{i}}$ and V$_{\textrm{Al}}$ were both found to have deep acceptor levels, with the average O$_{\textrm{i}}$ (0/-2) charge transfer level lying 3.4 eV below the a-Al$_2$O$_3$ CBM, and the average V$_{\textrm{Al}}$ (-2/-3) charge transfer level lying 3.5 eV below the CBM. However, their lack of occupied energy levels in the band gap means that they are not responsible for the transitions seen in the EPDS and GS-TSCIS measurements~\cite{Zahid2010}.

To explain spectral distribution of electron transitions observed experimentally~\cite{Zahid2010} (see Fig. \ref{fig:PDS}) a mechanism is proposed whereby the negatively charged O$_{\textrm{i}}$ and V$_{\textrm{Al}}$ states are compensated by the positively charged H$_{\textrm{i}}$, V$_{\textrm{O}}$ and Al$_{\textrm{i}}$ defects. As a result of the compensation, the as grown system has a relatively small net overall charge. Following electron injection, the states of Al$_{\textrm{i}}$, V$_{\textrm{O}}$ or H$_{\textrm{i}}$ in the band gap become occupied by electrons and subsequently optical transitions into the oxide conduction band can be observed. 

The hypothesis regarding the co-presence of oppositely charged states in the metal oxide has earlier been invoked on the basis of electron capture cross-section arguments~\cite{Afanasev2004}. The present study provides an atomistic picture of this peculiar electrostatic condition. Furthermore, supporting the proposed interpretations, the K-S energy levels of these defects, determined with respect to the amorphous alumina CBM (see Fig. \ref{fig:KSlevels}), overlap with the results of EPDS and GS-TSCIS measurements~\cite{Zahid2010}. Finally, it is worth noting that the attractive Coulomb potential of the electron trapping sites would promote filling of these traps even in the absence of an externally applied electric field, provided electrons have sufficient energy to tunnel to the traps. This additional electron trapping would be consistent with the observed increase of the net negative charge upon annealing of the Si/alumina structures \cite{Kuhnhold_SiAlO_APL2016}. All in one, the available experimental observations appear to be consistent with the results of our defect simulations.

\section{ACKNOWLEDGMENTS} 

OAD acknowledges Argonne National Laboratory, USA for financial support. ALS and JC acknowledge funding provided by the UK Engineering and Physical Sciences Research Council (EPSRC) under grants No. EP/K01739X/1 and EP/P013503/1 and by the Leverhulme Trust RPG-2016-135. Computer facilities on the Archer service have been provided via the UKs HPC Materials Chemistry Consortium (EPSRC Grant No. EP/L000202). The authors wish to thank A.-M. El-Sayed, D. Z. Gao and J. Strand for helpful discussions. 

\bibliographystyle{apsrev}
\bibliography{al2o3_defects}
\expandafter\ifx\csname natexlab\endcsname\relax\def\natexlab#1{#1}\fi
\expandafter\ifx\csname bibnamefont\endcsname\relax
  \def\bibnamefont#1{#1}\fi
\expandafter\ifx\csname bibfnamefont\endcsname\relax
  \def\bibfnamefont#1{#1}\fi
\expandafter\ifx\csname citenamefont\endcsname\relax
  \def\citenamefont#1{#1}\fi
\expandafter\ifx\csname url\endcsname\relax
  \def\url#1{\texttt{#1}}\fi
\expandafter\ifx\csname urlprefix\endcsname\relax\def\urlprefix{URL }\fi
\providecommand{\bibinfo}[2]{#2}
\providecommand{\eprint}[2][]{\url{#2}}

\end{document}